\documentclass[12pt]{article}
\pdfoutput=1
\usepackage{latexsym, bm, amssymb}
\usepackage{amsmath, amsfonts, amsthm}
\usepackage[usenames]{color}
\usepackage{graphicx, subfigure, float, caption}
\usepackage{enumitem}

\usepackage{booktabs, multirow, threeparttable}
\usepackage{wallpaper}
\usepackage{appendix}
\usepackage{dcolumn}
\usepackage[colorlinks=false]{hyperref}
\usepackage{natbib}
\usepackage{array}
\usepackage{algorithm, algorithmic}

\graphicspath{{./figure/}}

\newcolumntype{d}[1]{D{.}{.}{#1}}

\usepackage[markup=default]{changes}

\newtheorem{innercustomgeneric}{\customgenericname}
\providecommand{\customgenericname}{}
\newcommand{\newcustomtheorem}[2]{%
  \newenvironment{#1}[1]
  {%
   \renewcommand\customgenericname{#2}%
   \renewcommand\theinnercustomgeneric{##1}%
   \innercustomgeneric
  }
  {\endinnercustomgeneric}
}

\newcustomtheorem{customthm}{Theorem}
\newcustomtheorem{customlemma}{Lemma}

\begin{document}

\baselineskip=24pt

\pdfoutput=1
\begin{titlepage}
\vspace*{4cm}
\begin{center}
    \Large{Impact of Global Warming on Extreme Rainfall in Taiwan}
    
    \vspace{2cm}
    Cheng-Ching Lin
    
    \vspace{0.5cm}
    Master Thesis
    
    \vspace{0.5cm}
    National Tsing Hua University
    
    \vspace{0.5cm}
    2023
\end{center}
\end{titlepage}




%
%
%
%
%


\begin{center}
\textbf{\Large Abstract} \\
\end{center}

The relationship between global warming and extreme rainfalls in Taiwan was examined in this study. Taiwan rainfall data from TCCIP, a project led by MOST, were analyzed. North Hemisphere reference temperature data from NCEI led by NOAA. The yearly maximum of daily rainfall was focused on and the PGEV model, as proposed by Olafsdottir et al. \citep{olafsdottir2021extreme}, was used to fit the extreme values and make inferences. The PGEV model integrates the General Extreme Value (GEV) and Peak over Threshold (PoT) approaches, which are commonly used to analyze extreme data. Relative intensity and return value were used to show the connection between temperature and extreme rainfall.

Results indicated that the intensity of extreme rainfall in Taiwan increases as the temperature rises. However, the effects of global warming on the frequency and intensity of extreme rainfalls varied by region. In the north and south regions, the frequency of extreme rainfalls changed, while in the center and east regions, the intensity of extreme rainfalls changed. Furthermore, according to the return value analysis, extreme rainfalls are likely to occur more frequently in the future.

To account for differences between locations, Gaussian Process was used to smooth the results obtained using the PGEV model. In addition, simulations using the Gaussian copula and Gaussian Process were conducted to determine the quantile confidence intervals for each PGEV model. The simulations showed that all tests comparing with models with and without covariates are significant.  

\bigskip 


\addcontentsline{toc}{section}{Abstract}
\clearpage




%
\pagenumbering{arabic}


\section{Introduction}

Extreme value analysis draw more attention recently, especially in the fields of environment and finance. In finance, how to predict and prevent the damage of extreme financial risk is the main interest of risk management. Extreme value analysis is mainly used in quantifying tail distribution, finding the measurement of risks such as Value-at-Risk (VaR) and Expected Shortfall (ES), and describing the tails properties in the markets of stocks (\cite{nortey2015extreme}, \cite{sheraz2021extreme}), futures \citep{cotter2006extreme}, and cryptocurrencies \citep{gkillas2018application}. Nolde \& Zhou \citep{nolde2021extreme} provided a comprehensive introduction to the application of extreme value analysis in finance. In environmental science, extreme value analysis is applied to characterize the occurrence and intensity of natural disasters caused by extreme weather, such as floods and earthquakes. Extreme value analysis on spatial data is a popular topic in environmental research. Blanchet et al. \citep{blanchet2011spatial} modeled the extreme snow depth in Switzerland and found the directional effect of snowfall. Davison et al. \citep{davison2012statistical} introduced the classical modeling of spatial extremes, and Huser \& Wadsworth \citep{huser2022advances} provided a more advanced introduction to spatial extremes analysis.

There is two popular approaches in extreme value analysis. One is called the block maximum approach, and it focuses on maximum (or minimum) data in each time blocking, such as the daily maximum rainfall. The other is called the threshold exceedance approach, and it focuses on the data higher than a certain threshold. For modeling, the block maximum approach uses the generalized extreme value (GEV) model to characterize the tail distribution and probability; the threshold exceedance approach uses the Peak over Threshold (PoT) model which explores the intensity and frequency of the extreme events of interest.  
More details on extreme value models refer to the textbooks \citep{coles2001introduction} and review papers \citep{davison2015statistics}, which give an overall classical introduction to the concepts of extremes analysis.

According to the AR6 report (Sixth Assessment Report Climate Change 2021) from IPCC (Intergovernmental Panel
on Climate Change) \citep{RN1}, the global temperature is getting higher every year, and the global temperature is already risen by $1.1^{\circ}$C in 2021 relative to the period 1850-1900. 
Being aware of such climate change, Olafsdottir et al. \citep{olafsdottir2021extreme} aims to answer the question that the frequency and intensity of extreme rainfall changes as temperature rises in the Northeastern United States. To use the annual daily maximum precipitations with consistent data quality, and to maintain the analysis ability of the threshold exceedance approach, they combined GEV and PoT approaches with Poisson distribution to build a new model called PGEV to solve the problem. Using different settings in PGEV models, they analyze the impact on extreme rainfall under various scenarios of global warming. With the data evidence, they found that,
 as the temperature rises, the intensity of extreme rainfall is fairly stable but extreme rainfall events happen more often in the Northeastern United States. 

Similar to \cite{olafsdottir2021extreme} , this thesis try to answer the following question: 
\begin{center}
\it{Whether is frequency or intensity of extreme rainfall events changes \\
as the temperature becomes higher in Taiwan? }
\end{center}
\noindent To answer the question, this study analyzes Taiwan rainfall data sourced from the TCCIP (Taiwan Climate Change Estimation Information and Adaptation Knowledge Platform) of the MOST (Ministry of Science and Technology). The mission of the TCCIP project is to provide the climate change scientific data service to achieve the following goals, including strengthening the high-resolution model simulation ability in Taiwan, supporting local impact research, integrating climate adaptation services such as climate scenarios, risk information, and adaptation tools, and developing integration of adaptation knowledge and implementation framework to expand the application possibilities of science-based solutions.
Among the research in TCCIP, two studies are relevant to extreme rainfall analysis in Taiwan. Tung et al. \citep{tung2016evaluating} transformed daily precipitation into probability indices with cumulative distribution function of GEV distribution and use probability indices to detect whether the probability of extreme rainfall events becomes higher when years go by. 
Henny et al. \citep{henny2021extreme} used TCCIP gridded spatial rainfall data and defined five rainfall seasons to proceed the extreme value analysis under different rainfall types separately.  They also defined ``extreme rainfall events'' as the rainfall which is above the 99th percentile in each season and each location, and  use such extreme rainfall data to fit the Theil-Sen slope. Their result reveals that, in Winter, Spring, and Typhoon seasons, the frequency of extreme rainfall events in Taiwan becomes higher;  the intensity of extreme rainfall events also becomes stronger in Typhoon season.

This thesis adopts the PGEV approach \citep{olafsdottir2021extreme} to infer the relationship between temperature and the intensity and frequency of extreme rainfall in Taiwan. In particular, the temperature is used as a temporal covariate. Based on the fitted PGEV model, the analysis of relative frequency and return level of the extreme rainfall in Taiwan shall be investigated under some scenarios of temperature raise. In the end, the Gaussian process is used to smooth the PGEV parameters over the spatial region of interest for better visualization of parameter estimates and to quantify the uncertainty for modeling inference.



This thesis is organized as follows: Section 2 introduces the temperature and rainfall data and the exploratory data analysis. The extreme value models are introduced in Section 3. The parameter estimation and detail fitting procedures are given in Section 4. Section 5 presents the application of extreme rainfall in Taiwan. Section 6 concludes the thesis contribution. 
The appendix provides the simulation and the confidence interval construction for some inferences in the application using the Gaussian process.

\clearpage


\section{Data}

This work uses temperature in NCEI and annual maxima of daily precipitation data in TCCIP to process the following analysis. 

\subsection{Rainfall Data}

\begin{figure}[t]
\centering \small
\includegraphics[width=13cm]{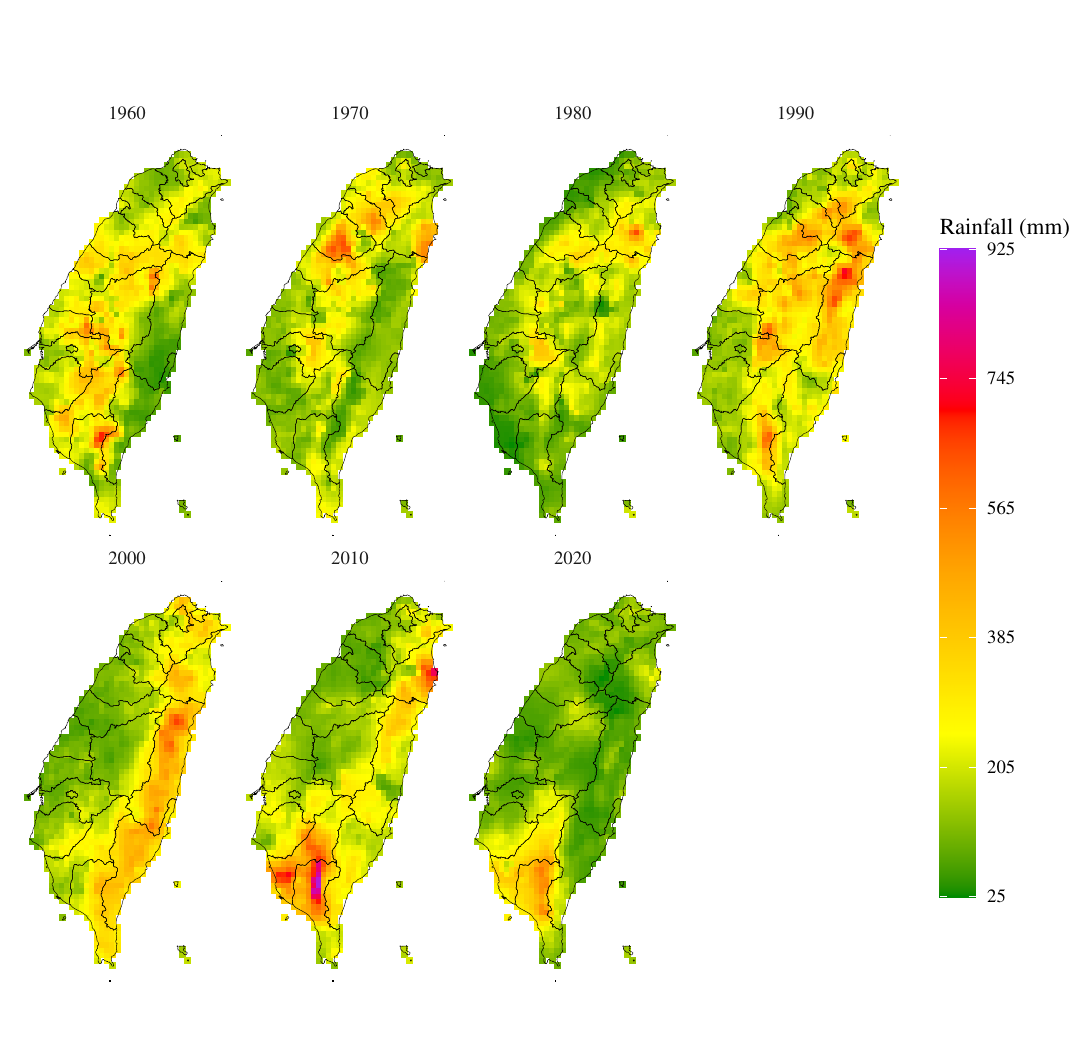}
\caption[Taiwan daily maximum rainfall plot from 1960 to 2020]{Annual maximum of daily rainfall in 5km $\times$ 5km resolution grid from TCCIP database in the period of 1960-2020.}
\label{Fig:Taiwan rainfall}
\end{figure}

\begin{figure}[t]
\centering \small
\begin{tabular}{cc}
\subfigure[The map of regions and boundary pixels]{\includegraphics[width=8cm]{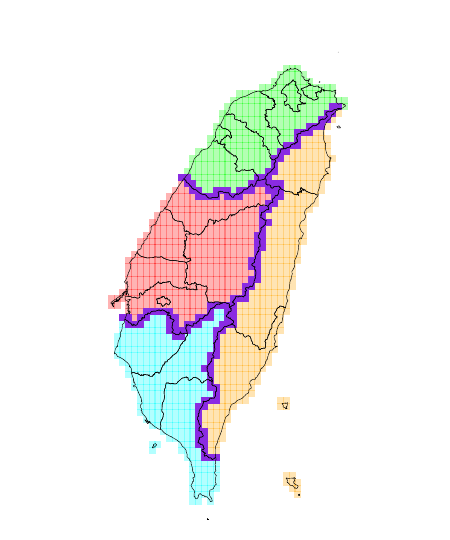}}&
\subfigure[The map shows the potential outliers pixels]{\includegraphics[width=8cm]{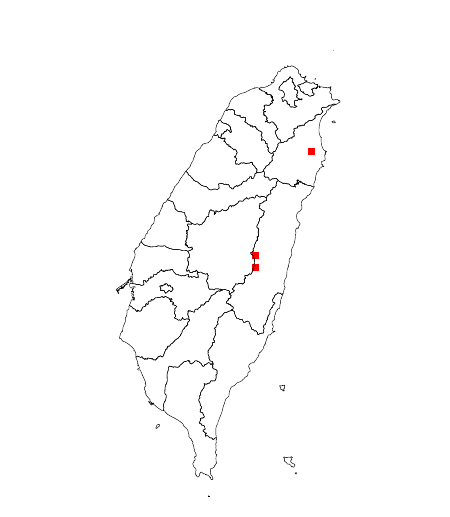}}
\end{tabular}
\caption[The pixel maps of Taiwan]{The left map shows the regions and boundary pixels. The green, red, cyan, and orange pixels represent North, Center, South, and East regions respectively. The violet pixels represent the overlapping boundaries of 4 regions in Taiwan; The right map shows the potential outliers pixels detected in Fig \ref{Fig:yearly median rainfall region}.} 
\label{Fig:Boundary repeated measurement}
\end{figure}

The precipitation data in TCCIP is jointly created by the Meteorological Bureau, the Water Resources Administration, the Civil Aviation Administration, the Environmental Protection Administration, and the Forestry Research Institute, among other units. The daily rainfall data are available from 1960 to 2020, on a 5-kilometer resolution grid. There are 1311 5km $\times$ 5km pixels covered Taiwan \citep{TCCIP2023}. 
The annual maximum of daily rainfall is extracted at each grid pixel and coupled with the latitude and longitude corresponding to the pixel. The annual maximum of daily precipitation in Taiwan during the period of 1960-2020 are shown in Fig \ref{Fig:Taiwan rainfall}. 
The TCCIP rainfall data are generated separately in 4 regions: North, Center, South, and East. Since the boundary of these regions is overlapped, there are replicated measurements that occurred at boundary pixels (violet pixels) shown in left map of Fig \ref{Fig:Boundary repeated measurement}. To handle this problem, this study only keeps the maximum among replicated measurements at each boundary location, since the underlying analysis focuses on extreme rainfall.  

\begin{figure}
\centering \small
\includegraphics[width=16cm]{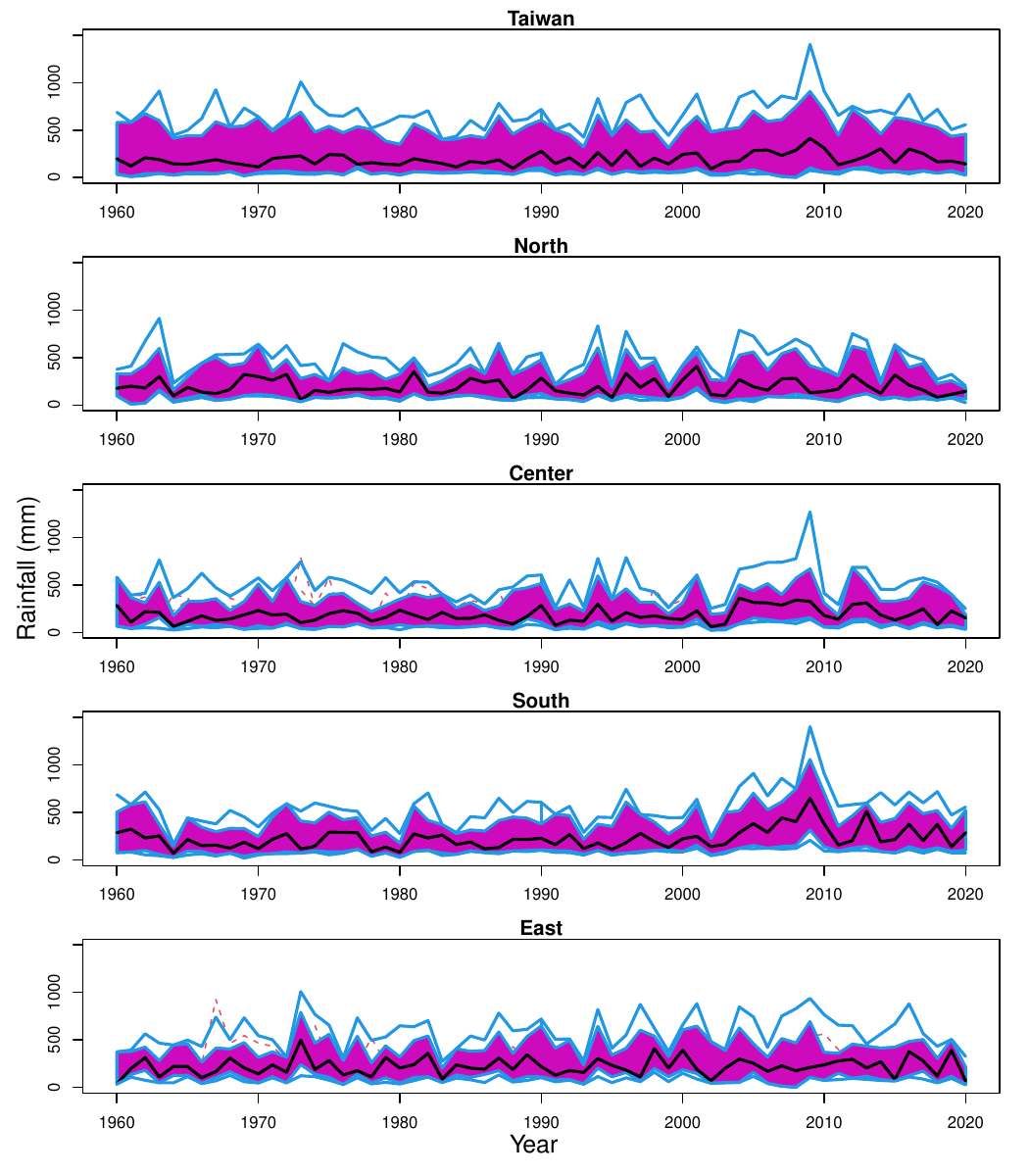}
\caption[The functional boxplot for different regions of Taiwan annual daily maximum rainfall from 1960 to 2020]{The functional boxplot of different regions of Taiwan's and whole Taiwan regions annual daily maximum rainfall in the period of 1960-2020.}
\label{Fig:yearly median rainfall region}
\end{figure}

To explore data patterns, $z_t(\bm{s}_j)$ at each location $\bm{s}_j$ is first viewed as a function of $t$. To visualize the temporal patterns of extreme rainfall events, the functional boxplot \citep{sun2011functional} of the annual maximum daily rainfall among all pixels and that among pixels in each subregion is shown in Fig \ref{Fig:yearly median rainfall region}. The x-axis indicates years, and the y-axis indicates the rainfall measurements (mm). Here, the range of y-axes in all subplolts are [0,1500]. The black line in the middle represents the median function. The pink highlights the region of 25\% to 75\% quantiles among the functions at all pixels. The upper and lower blue curves indicate 1.5 times of pink area height. If at least one point in a curve is outside the upper or lower blue curve, the curve is detected as a potential outlier. From Fig \ref{Fig:yearly median rainfall region}, the median of annual maximum daily rainfall does not have clear decreasing or increasing trend patterns for the entire or local regions. 
Generally speaking, the variation of annual maximum daily rainfall becomes larger and unstable in recent years. To see which region of Taiwan variates the most, Fig \ref{Fig:yearly median rainfall region} shows the functional boxplot of different regions. In the South region, a high peak occurred in 2009, due to typhoon Morakot, the deadliest typhoon in recorded history in Taiwan, hit the South region that year. In the Center region, the upper blue bound is far away from the center range (pink region) in the period of year 2000-2010, which means most of the locations in the Center regions have a large variation than before. Finally, in the North and East regions, the trend and variation are quite stable than in the Center and South regions.
There are three outlier candidates shown in red dashed lines. These outlier candidates are shown in right map of Fig \ref{Fig:Boundary repeated measurement} as red pixels. Two of candidates in Center region are at the boundary pixels, and the other one in East region is in the middle of Yilan.

\begin{figure}[t]
\centering \small
\includegraphics[width=13cm]{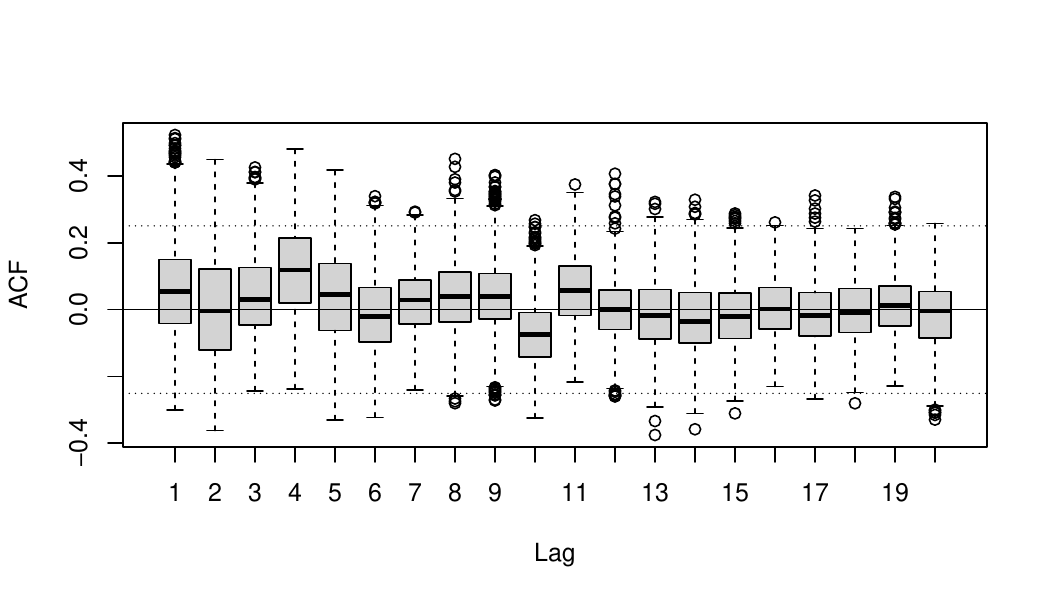}
\caption[The ACF of rainfall in Taiwan]{The x-axis means time lag, the y-axis means ACF of rainfall, and a box means the distribution of ACF at a certain lag in all pixels in Taiwan. For each lag, there are 1311 ACF values to construct a box. The dashed lines indicate the pointwise 95\% confidence bands. The solid line means 0.}
\label{Fig:acf_box}
\end{figure}

Finally, the temporal dependence in the series of annual maximum daily rainfall is checked. The sample ACFs are calculated for data at each pixel, and plotted as boxplot among all pixels for each lag in Fig \ref{Fig:acf_box}. Since no box is situated outside the confidence band (shown as the dotted lines), the annual maximum daily rainfall series can be assumed as temporal uncorrelated. Namely, later in the modeling and analysis, the correlations between rainfall observations among different years can be ignored.

\subsection{Temperature Data}


The thesis uses the North Hemisphere annual average reference temperature data provided by the National Centers for Environmental Information (NCEI), supported by the National Oceanic and Atmospheric Administration (NOAA). NCEI is the leading authority on environmental data. They maintain one of the largest archives of environmental research around the world. The temperature data refers to the average temperature during 1901-2000.

\begin{figure}[t]
\centering \small 
{\includegraphics[width=13cm]{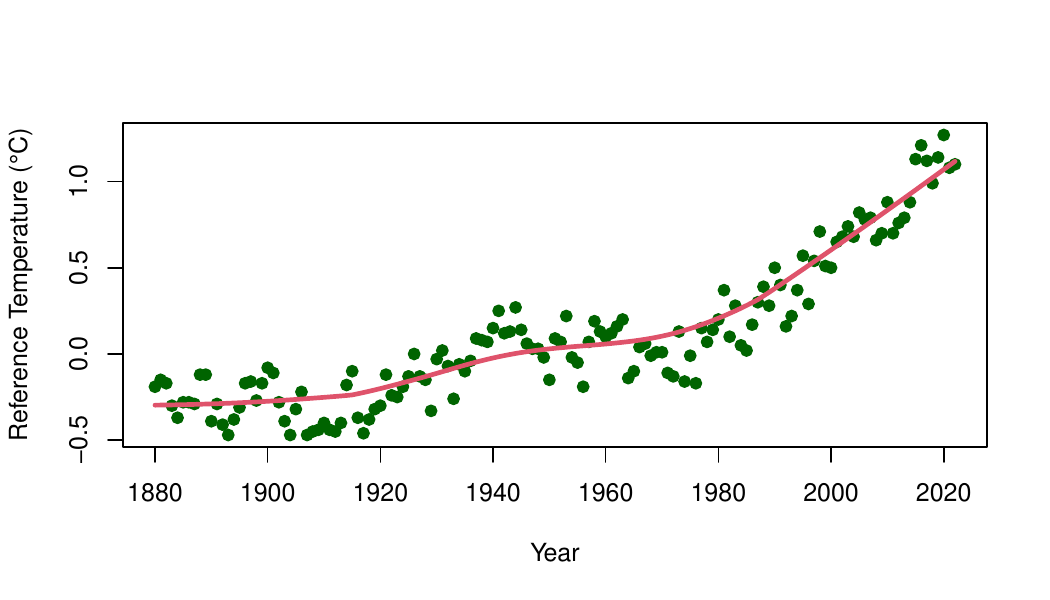}}
\caption[The average reference temperature in north hemisphere every year]{For each point is the annual average reference temperature in north hemisphere with respect to 1901-2000. The x-axis is years, and the y-axis is the reference temperature($^{\circ}$C). The red line is the smoothed reference temperature trend using Lowess method.} 
\label{Fig:Temperature}
\end{figure}

To explore the long-term trend patterns, the yearly reference temperature are further smoothed based on the Lowess method (Locally Weighted Scatterplot Smoothing, \cite{cleveland1979robust}), shown as the red line in Figure \ref{Fig:Temperature}. 
This ``smoothed temperature'' is considered as a covariate to study the effect of climate change in extreme rainfall analysis.

\subsection{Notations}

Finally, define the notations for data:
\begin{itemize}
\item each pixel in the grid is represented as $j$
\item latitude and longitude of each pixel are represented as a coordinate vector $\bm{s}_j$
\item $t$ represents year
\item annual maximum daily rainfall at location $\bm{s}_j$ and year $t$ is written as $Z_t(\bm{s}_j)$
\item the vector of observed yearly maximum of daily rainfall at location $\bm{s}_j$ from year 1 to $T$ is $\bm{z}(\bm{s}_j)\equiv\left(z_1(\bm{s}_j),\dots,z_T(\bm{s}_j)\right)'$
\item the smoothed reference temperature at year $t$ is denoted as $x_t$
\end{itemize}

\clearpage

\section{Models for Extreme Value Analysis}

In this section, two fundamental models in extreme value analysis, GEV and PoT, are introduced. The PGEV model \citep{olafsdottir2021extreme}, which integrates the formulation of both GEV and PoT models, will be introduced afterward.

\subsection{Generalized Extreme Value Models (GEV)}

For the GEV model (Coles, 2001, chapter 3), the data it fits are block maximum data. The maximum statistic among the observations in a block of time, such as month or year, follows the GEV model asymptotically. The CDF of GEV($\mu$, $\sigma$, $\gamma$) distribution satisfies
\begin{align} \label{eq:GEV}
G(z;\mu,\sigma,\gamma)=\exp\left[-\left(1+\gamma\cfrac{z-\mu}{\sigma}\right)^{-1/\gamma}\right], \quad \gamma\ne 0,
\end{align}
for $z$ satisfying $\gamma\left(z-\mu\right)<\sigma$,
where $\mu\in\mathbb{R}$ is the location parameter, $\sigma > 0$ is the  scale parameter, and $\gamma\in\mathbb{R}$ is the  shape parameter. The corresponding density of GEV($\mu$, $\sigma$, $\gamma$) is 
\begin{align}\label{eq:GEV_pdf}
f_G(z;\mu,\sigma,\gamma)=\cfrac{1}{\sigma}\left(1+\gamma\cfrac{z-\mu}{\sigma}\right)^{-1/\gamma-1}\exp\left[-\left(1+\gamma\cfrac{z-\mu}{\sigma}\right)^{-1/\gamma}\right], 
\end{align}
for $z$ satisfying $\gamma\left(z-\mu\right)<\sigma$.

If the shape parameters $\gamma=0$, the GEV model becomes a Gumbel model: 
\begin{align} \label{eq:GEV_nogamma}
G(z;\mu,\sigma)=\exp\left[-\exp\left(-\cfrac{z-\mu}{\sigma}\right)\right],\quad z\in\mathbb{R}.
\end{align}

The parameters in the GEV model have a close connection to those in the PoT model. Their connections are given in Section 3.3.

\subsection{Peak Over Threshold Models (PoT)}

For the PoT model (see for example Coles, 2001, chapter 4), the data it fit are the exceedance values over a high threshold $c$. The occurance of such over-threshold events follows a Poisson process with the rate parameter $\lambda_c$ depending on $c$. The exceedances are assumed to be mutually independent, independent of the Poisson process, and follow the PoT($\sigma_c$, $\gamma$) model. In particular, the exceedance $(Z_d-c)$ follows the generalized Pareto distribution with the CDF:
\begin{align} \label{eq:POT}
H_d(z;\sigma_c,\gamma) \equiv P(Z_d< c+z|Z_d>c)  =1-\left(1+\gamma\cfrac{z}{\sigma_c}\right)^{-1/\gamma},\quad\sigma_c+\gamma z>0,
\end{align}
\noindent
where $\sigma_c>0$ is a scale parameter depending on the threshold $c$, and $\gamma$ is a shape parameter. 
The corresponding density of $H_d(z;\sigma_c,\gamma)$ is 
\begin{align}\label{eq:POT_pdf}
f_H(z;\sigma_c,\gamma)=\cfrac{1}{\sigma_c}\left(1+\gamma\cfrac{z}{\sigma_c}\right)^{-1/\gamma-1},\quad\sigma_c+\gamma z>0.
\end{align}

If the shape parameter $\gamma=0$, the PoT model becomes an exponential model: 
\begin{align} \label{eq:POT_nogamma}
H_d(z;\sigma_c)=1-\exp\left(-\cfrac{z}{\sigma_c}\right),\quad z>0.
\end{align}

\subsection{PGEV Models}

This work aims to analyze the frequency and intensity of extreme rainfall events with annual daily maximum rainfall, but it cannot be done by the GEV model alone. Since the frequency and intensity of extreme rainfall events can be represented respectively as the rate of rainfall events above thresholds and the exceedences values from thresholds, these inference properties are came from PoT model not GEV model. Therefore, Olafsdottir et al. \citep{olafsdottir2021extreme} introduced the PGEV model, which fits the block maximum data but infers the parameters in PoT models. The logic is illustrated as follows. 

Suppose the extreme daily rainfall over a threshold $c$ followed by a Poisson process with a yearly rate $\lambda_{c}$. Let $N$ be the random number of excesses in a year and let $Z$ be the annual maximum of daily rainfall. It follows that
\begin{align} \label{eq:PGEV0}
P(Z<z) 
&=\sum_{n=0}^{365}P(Z<z|N=n)P(N=n) \notag\\
&\approx\sum_{n=0}^{\infty} \left[H_d(z-c;\sigma_c,\gamma)\right]^n\cfrac{\lambda_{c}^n}{n!} \, e^{-\lambda_{c}} \notag\\
&=\sum_{n=0}^{\infty}\cfrac{1}{n!}  \left\{\lambda_c\left[1-\left(1+\gamma\cfrac{z-c}{\sigma_c}\right)^{-1/\gamma}\right]\right\}^n \, e^{-\lambda_{c}}\notag\\
&=\exp\left\{\lambda_c\left[1-\left(1+\gamma\cfrac{z-c}{\sigma_c}\right)^{-1/\gamma}\right]\right\} \, e^{-\lambda_{c}}\notag\\&=\exp\left\{-\lambda_c\left[1+\gamma\cfrac{z-c}{\sigma_c}\right]^{-/\gamma}\right\}\notag\\&=\exp\left\{-\left[1+\lambda_c^{-\gamma}+\gamma\cfrac{z-c}{\sigma_c}-1\right]^{-1/\gamma}\right\}\notag\\&=\exp\left\{-\left[1+\cfrac{\gamma}{\lambda_c^\gamma\sigma_c}\left(\lambda_c^\gamma\left(z-c\right)+\cfrac{\sigma_c}{\gamma}-\cfrac{\lambda_c^\gamma\sigma_c}{\gamma}\right)\right]^{-1/\gamma}\right\}\notag\\&=\exp\left\{-\left[1+\gamma\cfrac{z-\left(c+\sigma_c\cfrac{\lambda_c^{\gamma}-1}{\gamma}\right)}{\sigma_c\lambda_c^{\gamma}}\right]^{-1/\gamma}\right\},
\end{align}
in which the third equation is due to \eqref{eq:POT}.
Olafsdottir et al. \citep{olafsdottir2021extreme} called \eqref{eq:PGEV0} the CDF of a PGEV($\sigma_c$, $\gamma_c$) model given the threshold $c$:
\begin{align} \label{eq:PGEV}
F_c(z) =\exp\left\{-\left[1+\gamma\cfrac{z-\left(c+\sigma_c\cfrac{\lambda_c^{\gamma}-1}{\gamma}\right)}{\sigma_c\lambda_c^{\gamma}}\right]^{-1/\gamma}\right\}.
\end{align}
The corresponding density of PGEV is
\begin{align} \label{eq:PGEV_pdf}
{\small
f_c(z) = \cfrac{1}{\sigma_c\lambda_c^\gamma}\left[1+\gamma\cfrac{z-\left(c+\sigma_c\cfrac{\lambda_c^{\gamma}-1}{\gamma}\right)}{\sigma_c\lambda_c^{\gamma}}\right]^{-1/\gamma-1}\exp\left\{-\left[1+\gamma\cfrac{z-\left(c+\sigma_c\cfrac{\lambda_c^{\gamma}-1}{\gamma}\right)}{\sigma_c\lambda_c^{\gamma}}\right]^{-1/\gamma}\right\}.
}
\end{align}

Compare \eqref{eq:PGEV} with \eqref{eq:GEV}, PGEV is a special case of GEV, i.e., GEV($\mu_{\mbox{\tiny GEV}}$, $\sigma_{\mbox{\tiny GEV}}$, $\gamma$), but parametrized via $(\lambda_c, \sigma_c, \gamma)$ given the threshold $c$, satisfying
\begin{align} \label{eq:compare_PGEV1}
\mu_{\mbox{\tiny GEV}} = c+\sigma_{c} \, \cfrac{\lambda_{c}^{\gamma}-1}{\gamma}, \quad \quad 
\sigma_{\mbox{\tiny GEV}}  =\sigma_{c} \, \lambda_{c}^{\gamma}.
\end{align}

To be more general, this thesis allows the covariate effect in the rate and scale parameters and considers the regression form for $(\lambda_c,\sigma_c)$ at each pixel $j$ and year $t$ as:
\begin{align} 
\log\lambda_{c}(\bm{s}_j, x_t) =\beta_{0j}+\beta_{1j}x_t,  \label{eq:nonstationary_POT_1} \\
\log\sigma_{c}(\bm{s}_j, x_t) =\alpha_{0j}+\alpha_{1j}x_t, \label{eq:nonstationary_POT_2}
\end{align}
where $x_t$ is the long-term temperature given in Section~2.2 shown in Fig \ref{Fig:Temperature}.

Consequently, \eqref{eq:compare_PGEV1} are allowed to be pixel-wise and year-specific forms:
\begin{align} \label{eq:compare_PGEV2}
\mu_{\mbox{\tiny GEV}}(\bm{s}_j,x_t) = c_j+\sigma_{c}(\bm{s}_j,x_t) \, \cfrac{\lambda_{c}(\bm{s}_j,x_t)^{\gamma_j}-1}{\gamma_j},\quad\quad 
\sigma_{\mbox{\tiny GEV}} (\bm{s}_j,x_t)=\sigma_{c}(\bm{s}_j,x_t) \, \lambda_{c}(\bm{s}_j,x_t)^{\gamma_j}.
\end{align}

\subsection{Temperature Effects on Intensity of Rainfall}

This work evaluates
the relative frequency change, relative scale change and quantifies the return level 
subject to temperature raises 
to investigate the impact of climite change on the distribution of the annual maximum daily rainfall in Taiwan.
The ideas about the relative frequency change and the return level quentification are 
adopted from the approach by Olafsdottir et al. (2021).

 \subsubsection{Relative Frequency and Scale Changes}
 
 According to \eqref{eq:nonstationary_POT_1} and \eqref{eq:nonstationary_POT_2}, the temperature effects the PGEV parameter $(\lambda_c, \sigma_c)$.
 Let $\Delta_x=x_{\mbox{\tiny high}}-x_{\mbox{\tiny low}}$ be the temperature change. 
The relative changes in the frequency parameter and the scale parameter at the pixel $j$ subject to temperature change $\Delta_x$ are
\begin{align}
\Delta_\lambda(\bm{s}_j) &\equiv \cfrac{\lambda_{c}(\bm{s}_j, x_{\mbox{\tiny high}})-\lambda_{c}(\bm{s}_j, x_{\mbox{\tiny low}})}{\lambda_{c}(\bm{s}_j, x_{\mbox{\tiny low}})}=\exp(\beta_{1j}\Delta_x)-1,  
 \label{eq:relative_intensity} \\
\Delta_\sigma(\bm{s}_j) &\equiv \cfrac{\sigma_{c}(\bm{s}_j, x_{\mbox{\tiny high}})-\sigma_{c}(\bm{s}_j, x_{\mbox{\tiny low}})}{\sigma_{c}(\bm{s}_j, x_{\mbox{\tiny low}})}=\exp(\alpha_{1j}\Delta_x)-1. 
 \label{eq:relative_scale}
\end{align}

Different scenarios of $\Delta_x$ will be discussed in the application.

\subsubsection{Return Levels}

A return level (denoted as $R_q$)  of the annual maximum daily rainfall with a return period of $1/q$ years is defined as a high threshold whose probability of exceedance is $q\in (0,1)$, i.e., the $(1-q)$-th quantile of the underlying extreme distribution. For example, for $q=0.05$, $R_{0.05}$ means the intensity could occurred once in a 20-year period.
For the GEV($\mu$, $\sigma$, $\gamma$) model \eqref{eq:GEV}, the return level satisfies
\begin{align}
R_q \equiv G^{-1}(1-q; \mu, \sigma,\gamma) = \mu-\cfrac{\sigma}{\gamma}\left\{\vphantom{\cfrac{\sigma}{\gamma}} 1-\left[-\log\left(1-q\right)\right]^{-\gamma}\right\},\quad 0< q< 1.
\end{align}

\noindent Plugging in the parameter estimates in Eq\eqref{eq:compare_PGEV2}, the return level under the temperature $x_t$ at pixel $j$ becomes
\begin{align*}
R_q(\bm{s}_j, x_t) &=\mu_{\mbox{\tiny GEV}}(\bm{s}_j,x_t)-\cfrac{\sigma_{\mbox{\tiny GEV}}(\bm{s}_j,x_t)}{\gamma_j}\{1-[-\log(1-q)]^{-\gamma_j}\}\\
&=\left[c_j+\cfrac{\sigma_c(\bm{s}_j,x_t)}{\gamma_j}(\lambda_c(\bm{s}_j,x_t)^{\gamma_j}-1)\right]-\cfrac{\lambda_c(\bm{s}_j,x_t)^{\gamma_j}\sigma_c(\bm{s}_j,x_t)}{\gamma_j}\{1-[-\log(1-q)]^{-\gamma_j}\}.
\end{align*}


To evaluate how the temperature raise affects the intensity of extreme events, the probability of a storm at pixel $j$ having the rainfall intensity exceeded over $R_q(\bm{s}_j, x_{\mbox{\tiny low}})$ subject to a temperature change $\Delta_x= x_{\mbox{\tiny high}} - x_{\mbox{\tiny low}}$ is defined as 
\begin{align} \label{eq: relative_return_level_change}
\begin{split}
& P\left[Z_t(\bm{s}_j)>R_q(\bm{s}_j, x_{\mbox{\tiny low}}) |  x_t=x_{\mbox{\tiny high}}\right] \\
& \quad =1-\exp\left\{-\left[1+\gamma_j\cfrac{R_q(\bm{s}_j,x_{\mbox{\tiny low}})-\left(c_j+\cfrac{\sigma_c(\bm{s}_j,x_{\mbox{\tiny high}})}{\gamma_j}\left[\lambda_c(\bm{s}_j,x_{\mbox{\tiny high}})^{\gamma_j}-1\right]\right)}{\lambda_c(\bm{s}_j,x_{\mbox{\tiny high}})^{\gamma_j}\sigma_c(\bm{s}_j,x_{\mbox{\tiny high}})}\right]^{-1/\gamma_j}\right\}
\end{split}
\end{align}

If covariate has no effects on parameters, i.e., $\beta_{1j}=\alpha_{1j}=0$,  Eq \eqref{eq: relative_return_level_change} reduces to $q$. 
If the  covariate only effects on the rate parameter, i.e., $\alpha_{1j}=0, \beta_{1j}\ne 0$, Eq \eqref{eq: relative_return_level_change} is simplified as
\begin{align} \label{eq:rate_return_level}
P\left[Z_t(\bm{s}_j)>R_q(\bm{s}_j, x_{\mbox{\tiny low}}) |  x_t=x_{\mbox{\tiny high}}\right] = 1-\exp\left[\cfrac{\lambda_c(\bm{s}_j, x_{\mbox{\tiny high}})}{\lambda_c(\bm{s}_j, x_{\mbox{\tiny low}})}\log\left(1-q\right)\right] = 1-(1-q)^{\Delta_\lambda(\bm{s}_j)+1}.
\end{align}  
\noindent 
In particular, Eq \eqref{eq:rate_return_level} only relies on the relative changes in frequency parameter $\lambda$ defined in Eq \eqref{eq:relative_intensity}.
On the other hand, if the  covariate only effects on the scale parameter, i.e., $\alpha_{1j}\ne 0, \beta_{1j}= 0$, Eq \eqref{eq: relative_return_level_change} is simplified as
\begin{align} \label{eq:scale_return_level}
& P\left[Z_t(\bm{s}_j)>R_q(\bm{s}_j, x_{\mbox{\tiny low}}) |  x_t=x_{\mbox{\tiny high}}\right] \notag \\
& \quad =1-\exp\left\{-\exp(\beta_{0j})\left[1-\cfrac{\left(1-\exp(-\beta_0j)\left(-\log(1-q)\right)^{-\gamma_j}\right)}{\Delta_{\sigma}(\bm{s}_j)+1}\right]^{-1/\gamma_j}\right\}.
\end{align}  
Eq \eqref{eq:scale_return_level} means that if the relative intensity changes by $\Delta_{\sigma}(\bm{s}_j)$, for temperature $\Delta_x$ increase, then a $R_q$-level storm will happen with Eq \eqref{eq: relative_return_level_change}. That is, the probability of $R_q$-level storms in $x_{\mbox{\tiny low}}$ increase under $x_{\mbox{\tiny high}}$.

\clearpage


\section{Estimation}

This study assumes the nonstationary PGEV model with spatial and temporal varying parameters 
$(\lambda_c(\bm{s}_j,x_t), \sigma_c(\bm{s}_j,x_t), \gamma_j)$ at year $t$ and pixel $j$ satisfying \eqref{eq:nonstationary_POT_1} and \eqref{eq:nonstationary_POT_2}. 

For simplicity, this work only considers model fitting at each pixel using the maximum likelihood estimation.

\subsection{Determining Threshold}

To fit the PGEV model, the threshold $c_j$ for daily rainfall of each pixel $j$ is needed to be determined in \eqref{eq:PGEV} in advance. For simplicity, this thesis determines the thresholds $c_j$ by fitting GEV with extreme value rainfall for each pixel $j$ and ignoring the effect of temperature. That is, find the MLE of parameters from pdf of GEV Eq \eqref{eq:PGEV_pdf}. The threshold is determined as follows that, an extreme rainfall event is defined as rainfall that only happens on average in about $365.25(1-p)$ days. The PGEV parameters in Eq \eqref{eq:compare_PGEV1} can be transformed into the formula below:

\begin{align} \label{eq:threshold}
c_j=\hat{\mu}_j - \cfrac{\hat{\sigma}_j(1-\lambda_p^{-\hat{\gamma}_j})}{\hat{\gamma}_j},
\end{align}
\noindent
where $\lambda_p = 365.25(1-p)$.

\subsection{Maximum Likelihood Estimation for PGEV}

The study use maximum likelihood (ML) to  estimate the parameters $\bm{\theta}_j = (\beta_{0j},\beta_{1j}, \alpha_{0j}, \alpha_{1j}, \gamma_j)'$ in PGEV model for each pixel $j$. Suppose the last temporal index is $T$. 
According to \eqref{eq:PGEV_pdf}, the log-likelihood of $\bm{\theta}_j$ given the observations
$\bm{z}(\bm{s}_j)\equiv (z_1(\bm{s}_j),\dots, z_T(\bm{s}_j))'$ at the $j$th pixel is 
\begin{align} \label{eq:PGEV_likelihood}
\begin{split}
& \ell(\bm{\theta}_j)  = \sum_{t=1}^T \log f_{c_j}(z_t(\bm{s}_j)) \\ 
& \quad = -\sum^{T}_{t=1}\left\{\left(1+\cfrac{1}{\gamma_j}\right)\log\left[\gamma_j\left(z_t(\bm{s}_j)-c_j\right)+\exp\left(\alpha_{0j}+\alpha_{1j}x_t\right)\right]\right\}\\
& \quad \quad -\sum^{T}_{t=1}\left\{\exp\left(\beta_{0j}+\beta_{1j}x_t\right)\left[\gamma_j\left(z_t(\bm{s}_j)-c_j\right)\exp\left(\alpha_{0j}+\alpha_{1j}x_t\right)+1\right]^{-1/\gamma_j}\right\}\\
& \quad \quad +\sum^{T}_{t=1}\left\{\cfrac{1}{\gamma_j}\left(\alpha_{0j}+\alpha_{1j}x_t\right)+\left(\beta_{0j}+\beta_{1j}x_t\right)\right\}.
\end{split}
\end{align}

\noindent 
The ML estimate of $\bm{\theta}_j$ is defined as
\begin{align}
\hat{\bm{\theta}}_j = \mbox{arg}\max \,  \ell(\bm{\theta}_j) ,
\end{align}
which is found using the maxLik package \citep{henningsen2011maxlik} in R. 

\subsection{Hypothesis Testing and Model Selection}

To find out whether the frequency or the intensity of extreme rainfall changes with temperature, this work fits four PGEV models as Eq \eqref{eq:PGEV} at each pixel $j$ and finds MLE of parameters by solving log-likelihood Eq \eqref{eq:PGEV_likelihood}. The four models are: 
\begin{itemize}
 \item $PGEV_{\lambda, \sigma}$ model: the frequency and scale parameters varying with temperature and the parameter space is 
 \begin{align*}
\Omega_{\lambda, \sigma} = \left\{\bm{\theta}:\beta_0\in\mathbb{R},\beta_1\in\mathbb{R}, \alpha_0\in\mathbb{R}, \alpha_1\in\mathbb{R}, \gamma\in\mathbb{R}\right\}.
\end{align*}
\noindent
The correspondence MLE is denoted as $\hat{\bm{\theta}}_j^{(\lambda, \sigma)} = \underset{\bm{\theta}_j\in\Omega}{\mbox{arg}\max} \,  \ell(\bm{\theta}_j)$

\item $PGEV_{\lambda}$ model: only the frequency parameter changes with temperature and the parameter space is 
\[
\Omega_\lambda = \Omega\cap\left\{\alpha_1=0\right\}.
\]
The correspondence MLE is $\hat{\bm{\theta}}_j^{(\lambda)} = \underset{\bm{\theta}_j\in\Omega_\lambda}{\mbox{arg}\max} \,  \ell(\bm{\theta}_j) .$

\item $PGEV_{\sigma}$ model: only the scale parameter changes with temperature and the parameter space is 
\[
\Omega_\sigma = \Omega\cap\left\{\beta_1=0\right\}.
\]
The correspondence MLE is $\hat{\bm{\theta}}_j^{(\sigma)} = \underset{\bm{\theta}_j\in\Omega_\sigma}{\mbox{arg}\max} \,  \ell(\bm{\theta}_j)$.

\item  $PGEV_0$ model: the parameters is time-invariant and the parameter space is 
\[
\Omega_0 = \Omega\cap\left\{\beta_1=\alpha_1=0\right\}.
\]
The correspondence MLE is $\hat{\bm{\theta}}_j^{(0)} = \underset{\bm{\theta}_j\in\Omega_0}{\mbox{arg}\max} \,  \ell(\bm{\theta}_j)$.
\end{itemize}

\medskip 

To select the appropriate model, this study conducts three Likelihood Ratio tests (LRT) and gets those p-values for each pixel $j$. The three tests are:

\begin{enumerate}
\item[] {\bf Test 1} \quad\quad $H_0$: $\bm{\theta}_j\in\Omega_0$ vs $H_1$: $\bm{\theta}_j\in\Omega_\lambda$,
\item[] {\bf Test 2} \quad\quad $H_0$: $\bm{\theta}_j\in\Omega_0$ vs $H_1$: $\bm{\theta}_j\in\Omega_\sigma$,
\item[] {\bf Test 3} \quad\quad  $H_0$: $\bm{\theta}_j\in\Omega_0$ vs $H_1$: $\bm{\theta}_j\in\Omega_{\lambda, \sigma}$.
\end{enumerate}

Accordingly, the LRT test statistics for Tests 1-3 and their null distribution under $H_0$ are given as follows respectively:

\begin{enumerate}
\item [] {\bf Test 1}  
\begin{align}\label{eq:Lambda_lambda}
\Lambda_\lambda=-2\left[\ell(\hat{\bm{\theta}}_j^{(0)})-\ell(\hat{\bm{\theta}}_j^{(\lambda)})\right]\sim\chi^2_1,
\end{align}
\item [] {\bf Test 2} 
\begin{align}\label{eq:Lambda_sigma}
\Lambda_\sigma=-2\left[\ell(\hat{\bm{\theta}}_j^{(0)})-\ell(\hat{\bm{\theta}}_j^{(\sigma)})\right]\sim\chi^2_1,
\end{align}
\item [] {\bf Test 3}  
\begin{align}\label{eq:Lambda}
\Lambda_{\lambda, \sigma}=-2\left[\ell(\hat{\bm{\theta}}_j^{(0)})-\ell(\hat{\bm{\theta}}_j^{(\lambda, \sigma)})\right]\sim\chi^2_2.
\end{align}
\end{enumerate}

Besides LRTs, AIC (Akaike's Information Criteria, \cite{akaike1974new}) is also used for model selection. 
Let $\hat{\bm{\theta}}$ be a generic notation for MLE of one of PGEV models. The definition of AIC is:
\begin{align}\label{eq:AIC}
AIC(\hat{\bm{\theta}}) = -2\ell(\hat{\bm{\theta}}) + 2\dim(\hat{\bm{\theta}}),
\end{align}
where $\dim(\hat{\bm{\theta}})$ denotes the number of parameters in $\hat{\bm{\theta}}$.
AIC will be calculated for the models fitted to each pixel. The lower value of AIC, the better fitting of models. The whole fitting procedure is outlined in Algorithm \ref{alg:MLE}.  

\begin{algorithm}[h]
  \caption{Steps for finding MLE of PGEV}
  \label{alg:MLE}
  \begin{algorithmic}[1]
  \REQUIRE
  $p$ for determining thresholds $c_j$;
  \FOR{each $j\in[1,1311]$}
    \STATE Fit GEV($\mu$,$\sigma$,$\gamma$) with annual maximum data $\bm{z}(\bm{s}_j)$ to get $\hat{\mu}_j$,$\hat{\sigma}_j$,$\hat{\gamma}_j$.
    \STATE Substitute the parameters from previous step into Eq \eqref{eq:threshold}, and determine the threshold $c_j$.
    \STATE Set $\hat{\mu}_j$,$\hat{\sigma}_j$,$\hat{\gamma}_j$ from step 2 as initial value parameters of PGEV model, and use BFGS and Eq \eqref{eq:PGEV_likelihood} to find the MLE $\hat{\bm{\theta}}_j$. (Note: If an error occurs, the initial value of the shape parameter $\gamma_j$ is changed to $10^{-8}$)
    \STATE For all models with covariates $x_t$, perform LRT on the stationary models, get the p-value for each model and calculate AIC for each model.
  \ENDFOR
  \end{algorithmic}
\end{algorithm}

\clearpage

\section{Application}
\subsection{GEV Parameters and Threshold}

This work uses the annual maximum daily rainfall data $\bm{z}_t(\bm{s}_j)$ from TCCIP to analysis the extreme rainfall property in Taiwan. The temperature data $x_t$ is from NCEI. The period of data is 61 years ($t=1,2,...,61$) from 1960 to 2020 and $j=1,2,...,1311$ in a 5km x 5km grid.

\begin{figure}
\centering \small
\begin{tabular}{lcr}
\subfigure[location parameters $\hat{\mu}_j$]{\includegraphics[width=4.5cm]{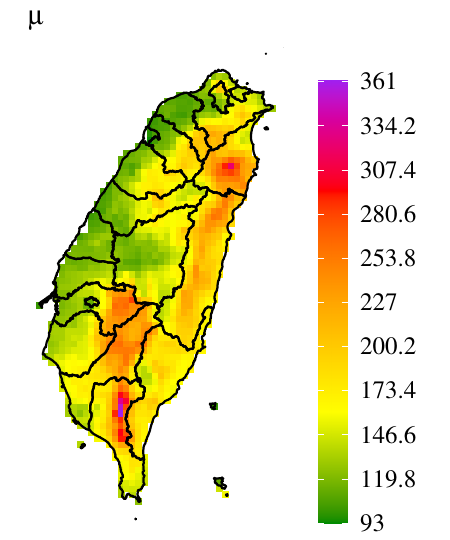}}&
\subfigure[scale parameters $\hat{\sigma}_j$]{\includegraphics[width=4.5cm]{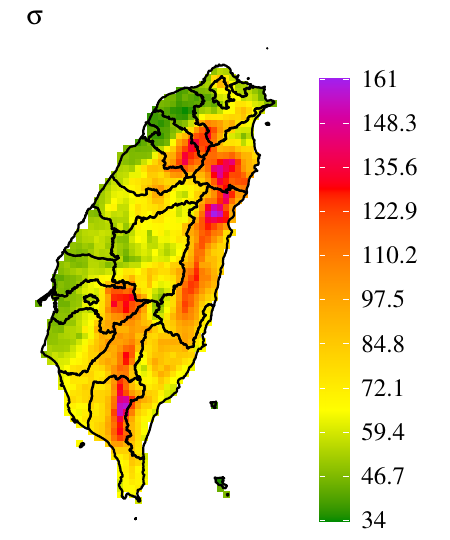}}&
\subfigure[shape parameters $\hat{\gamma}_j$]{\includegraphics[width=4cm]{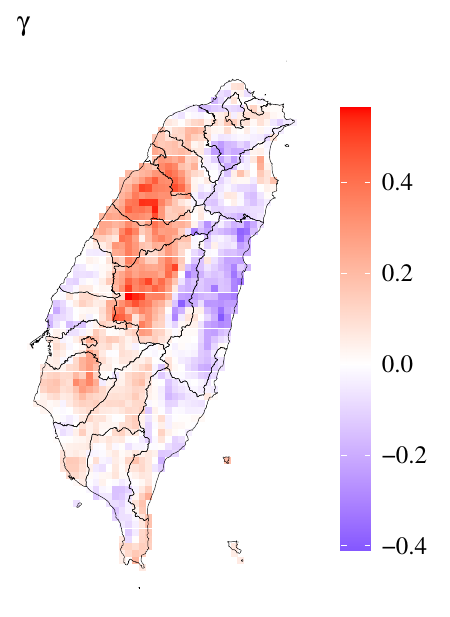}}
\end{tabular}
\caption[The map of GEV parameters]{Parameter estimates fitted by GEV. } 
\label{Fig:GEV_map}
\end{figure}

First, fit GEV($\mu_j$, $\sigma_j$, $\gamma_j$) models as Eq \eqref{eq:GEV} ignoring the effect of temperature for each pixel $j$. The estimated parameters of GEV are shown in Fig \ref{Fig:GEV_map} as maps. From Fig \ref{Fig:GEV_map}, the left map shows the location parameters $\mu_j$ on each pixel in Taiwan, the middle of the South area has larger values, indicating those areas have a larger yearly maximum of daily rainfalls; the middle map shows the scale parameters $\sigma_j$, the South and the East regions have larger values, showing the variation in these two regions are higher than other regions; the right map shows the shape parameters $\gamma_j$, have a clear difference between Eastern and Western of Taiwan.

\begin{figure}
\centering \small
\begin{tabular}{cc}
\subfigure[The histogram of thresholds]{\includegraphics[width=10cm]{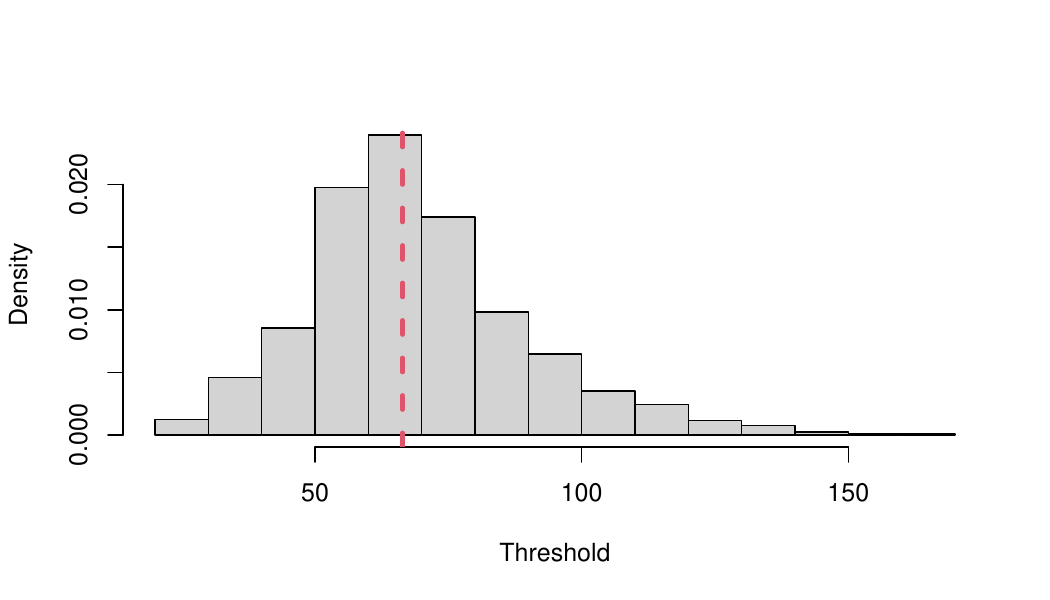}}&
\subfigure[The map of thresholds]{\includegraphics[width=6cm]{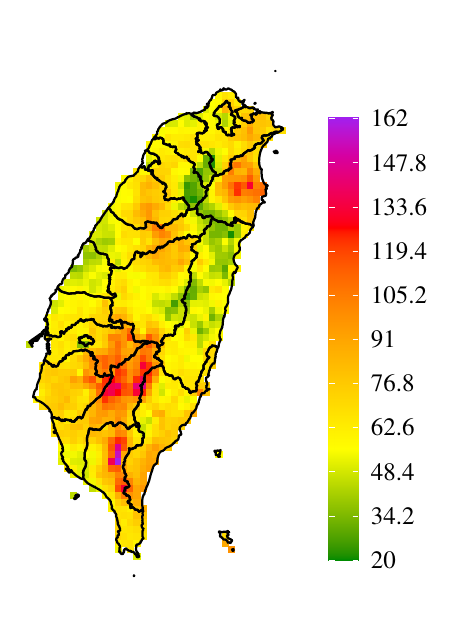}}
\end{tabular}
\caption[The plots for thresholds]{The histogram of thresholds is the left one, the dashed line is the median of thresholds. The right one is the map of thresholds.} 
\label{Fig:threshold}
\end{figure}

After getting the parameters from GEV models, the thresholds $c_j$ for PGEV models are determined
using Eq \eqref{Fig:threshold}. Here, the thresholds are determined by the quantile of $p = 0.99$ in a GEV stationary model. The distribution of thresholds among pixels is shown in Fig \ref{Fig:threshold}. The histogram of thresholds is right-skewed, that represents few places having large extreme rainfall only. According to the map on the right of Fig \ref{Fig:threshold}, the southeast area and Yilan have large thresholds, and this pattern is consistent with the rainfall maps from Fig \ref{Fig:Taiwan rainfall}.

\subsection{PGEV Fitting and Model Comparison}

\begin{figure}
\centering \small 
{\includegraphics[width=12cm]{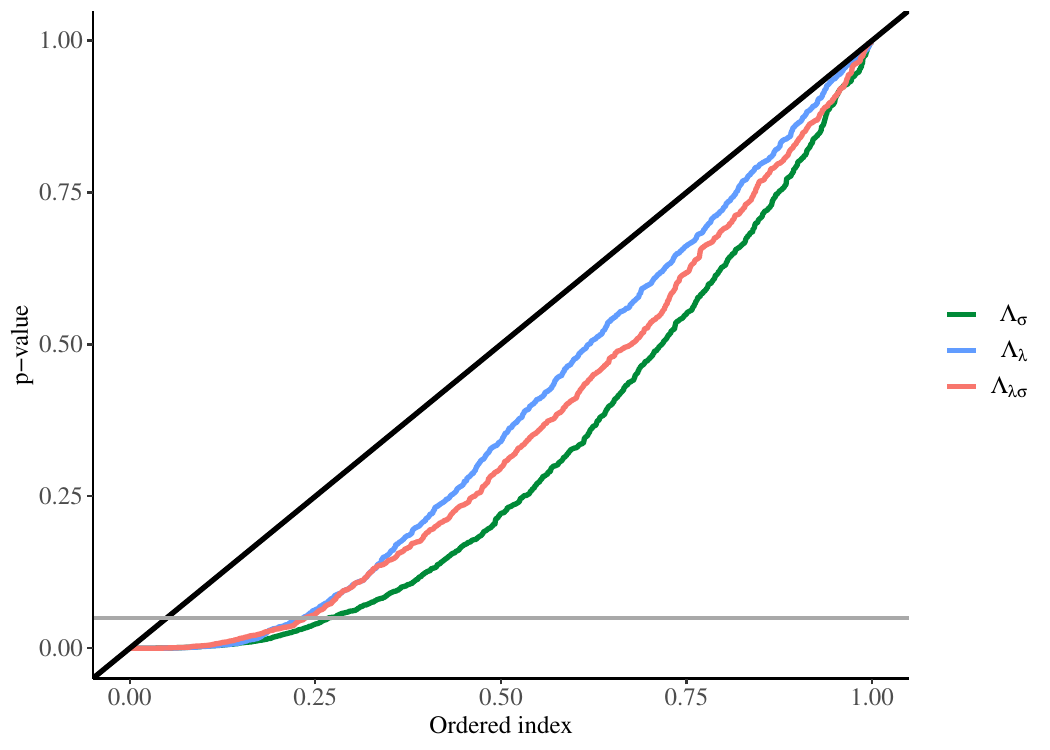}}
\caption[PGEV p-value QQplot]{This is the QQplot of the p-value from LRT. The x-axis is ordered index and normalize it into $[0,1]$. The black line is diagonal line which means not significant at all. The colored line means LRT p-value of each test. The gray line is 0.05. } 
\label{Fig:pvalue QQplot}
\end{figure}

\begin{figure}
\centering \small 
{\includegraphics[width=15cm]{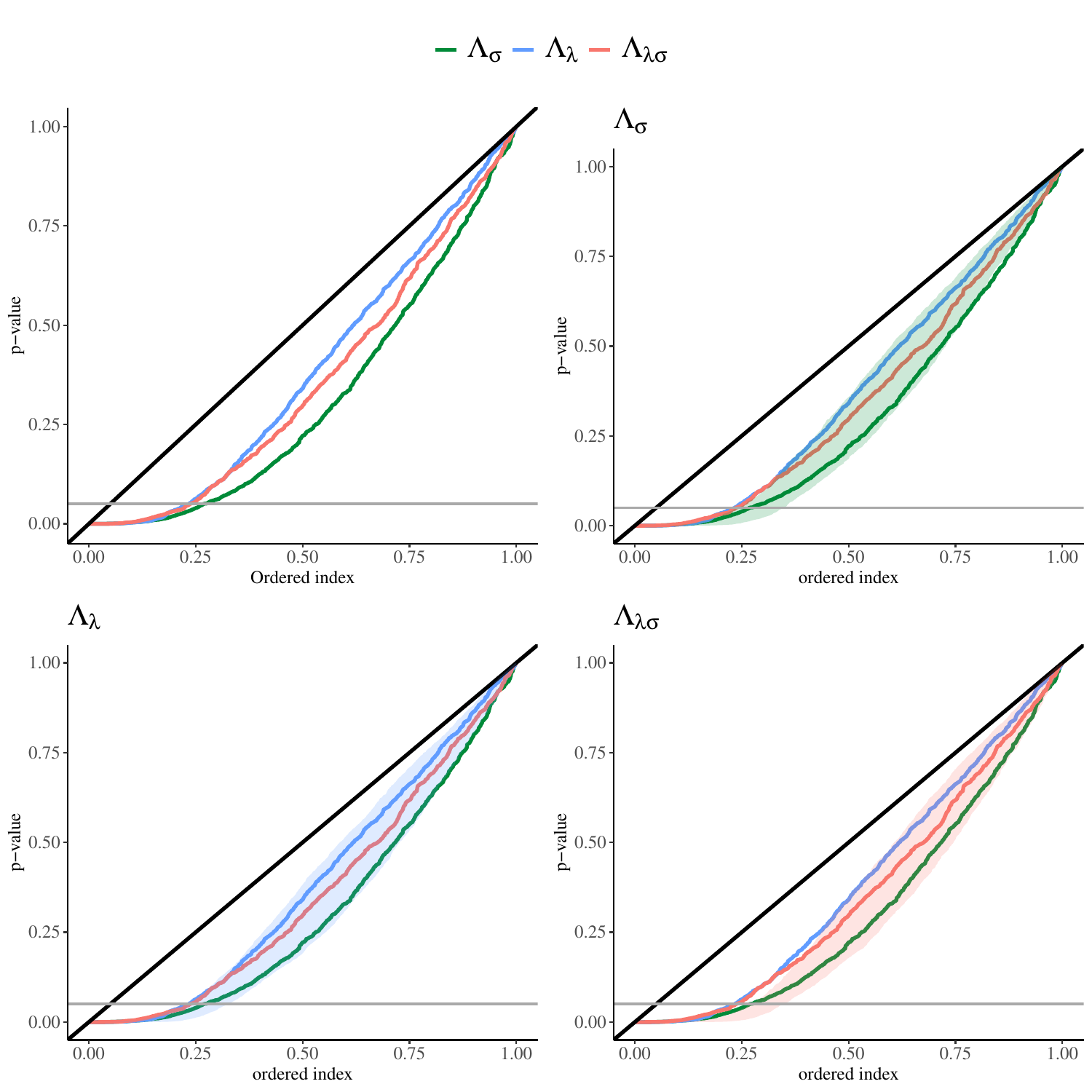}}
\caption[The PGEV p-value QQplot with confidence band]{The simulation result is the band in the picture above, and the previous lines are the same as seen in Fig \ref{Fig:pvalue QQplot}. The top left is the QQplot of  p-values, and the others are the QQplot of p-values with confidence band for each test.} 
\label{Fig:simulation MVN}
\end{figure}

After determining thresholds $c_j$, the four PGEV models are fitted with the data $\bm{z}(\bm{s}_j)$  for each pixel $j$ and the best model is determined by the test results of LRT $\Lambda_\lambda$, $\Lambda_\sigma$, and $\Lambda$ defined in Eq \eqref{eq:Lambda_lambda} -- Eq \eqref{eq:Lambda} respectively. These tests are conducted for all pixels in Taiwan. The p-values of each test are collected.
For each test, the p-values among pixels are sorted in ascending order, denoted as $p_{(i)}$, and plotted against $i/1311$ as a curve in Fig \ref{Fig:pvalue QQplot}. If one of the three tests is significant in most of the pixels in Taiwan, the representative curve will be concave-down and vice versa. From Fig \ref{Fig:pvalue QQplot}, $PGEV_{\sigma}$ is most significant compared to the other models since the curve of $\Lambda_\sigma$ is most concave-down. 
The confidence band of each QQ-plot curve is constructed via the bootstrap method. The detailed procedure is provided in Appendix II. The curves with the corresponding confidence band are shown in Fig \ref{Fig:simulation MVN}. From top left of Fig \ref{Fig:simulation MVN}, all lines are inside the confidence band. According to the confidence band, the p-values from three tests are not significantly superior to others.

\begin{figure}
\centering \small 
{\includegraphics[width=14cm]{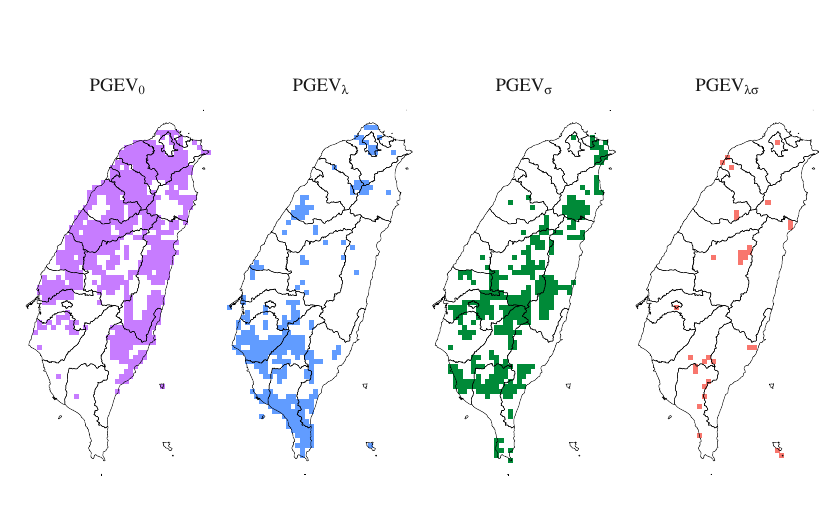}}
\caption[The map of pixels with lowest AIC models]{This is the map of pixels with lowest AIC models.} 
\label{Fig:AIC map}
\end{figure}

\begin{figure}
\centering \small 
{\includegraphics[width=12cm]{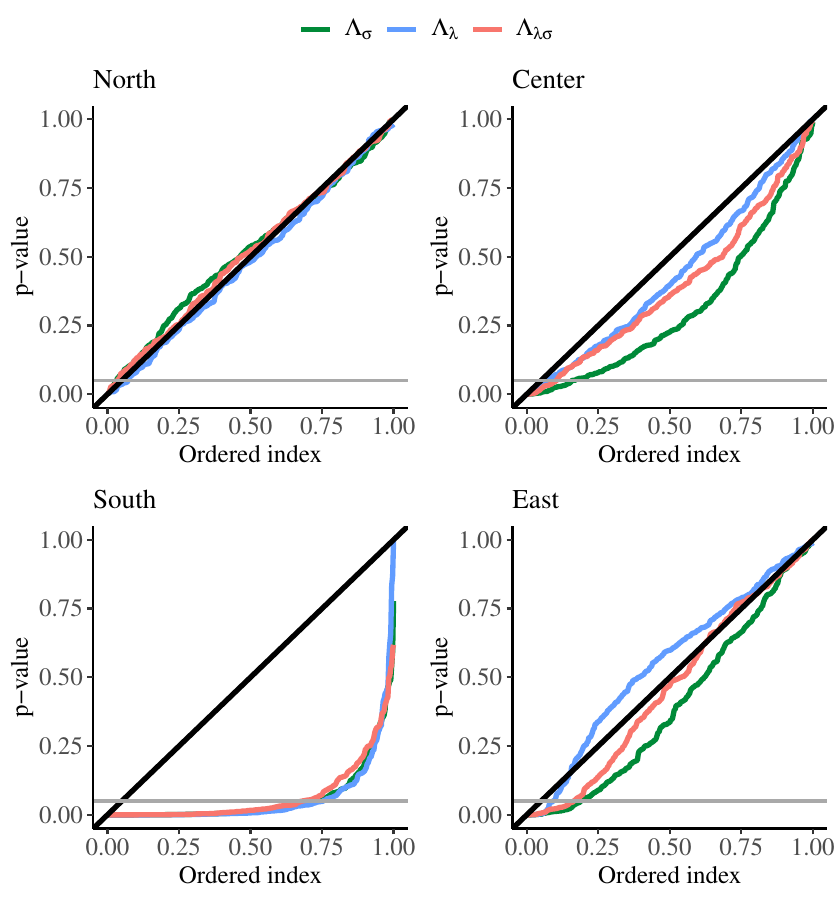}}
\caption[PGEV p-value QQplot separated into 4 regions]{This is the QQplot of the p-value from LRT but separate into 4 regions.} 
\label{Fig:pvalue QQplot region}
\end{figure}

Moreover, the best model can be selected by AIC defined in Eq \eqref{eq:AIC}. Since the original data come from 4 regions: North, Center, South, and East, the fitting performance of 4 fitted PGEV models in each region is summarized in Table \ref{tab:AIC Table Region}. Also, Table \ref{tab:AIC Table zero} put the $PGEV_0$ models into the comparison.
The frequency given in the table is the counts among total pixels in each subregion having the smallest AIC values among 4 fitted PGEV models. From Table \ref{tab:AIC Table Region}, $PGEV_\sigma$ model is the most favored model with the lowest AIC in the Center and East regions, while $PGEV_\lambda$ model is in the North and South regions. From Table \ref{tab:AIC Table zero}, $PGEV_0$ is most favored in all regions except south. In South region, $PGEV_\lambda$ is most favored model. However, $PGEV_\sigma$ is favored in more pixels than $PGEV_\lambda$. Fig \ref{Fig:AIC map} shows the pixels with lowest AIC models. Lots of pixels in the south regions are favored $PGEV_\lambda$, while $PGEV_\sigma$ is favored in Yilan and the center region. Moreover, those pixels favored $PGEV_\sigma$ are clustered as band across Taiwan from west to east.
Similarly, QQ plot on p-values for each subregion is shown in Fig \ref{Fig:pvalue QQplot region}. 
The p-value in the south is particularly significant compared to those in the other regions, while the north is the least significant region. 
The results show that extreme rainfalls are more likely to be affected by temperature, no matter freqency or intensity, in the South region.

\begin{table}
\caption{The frequency of the smallest AIC PGEV models in different regions}
\small
\centering 
\begin{tabular}{l|ccc|c}
\toprule[2pt]
Regions & $PGEV_{\lambda}$ & $PGEV_{\sigma}$ & $PGEV_{\lambda,\sigma}$ & Total pixels \\ \hline
North   & 141           & 99 & 15    & 255              \\
Center  & 141           & 281 & 12     & 434            \\
South   & 176           & 117          & 11     & 304             \\ 
East    & 101           & 208          & 9     & 318            \\ \hline
Total & 559 & 705 & 47 & 1311 \\
\bottomrule[2pt]
\end{tabular}
\label{tab:AIC Table Region}
\end{table}

\begin{table}
\caption{The frequency of the smallest AIC PGEV models in different regions with $PGEV_0$}
\small
\centering 
\begin{tabular}{l|cccc|c}
\toprule[2pt]
Regions & $PGEV_0$ & $PGEV_{\lambda}$ & $PGEV_{\sigma}$ & $PGEV_{\lambda,\sigma}$ & Total pixels \\ \hline
North & 198   & 34           & 18 & 5    & 255              \\
Center & 249  & 47           & 128 & 10     & 434            \\
South & 15   & 171           & 107          & 11     & 304             \\ 
East & 200    & 26           & 85          & 7     & 318            \\ \hline
Total & 662 & 278 & 338 & 33 & 1311 \\
\bottomrule[2pt]
\end{tabular}
\label{tab:AIC Table zero}
\end{table}

\subsection{Inference on Climate Change}

Since Fig \ref{Fig:pvalue QQplot region} and Table \ref{tab:AIC Table Region} shows that $PGEV_\sigma$ is significant in Center and East regions, while $PGEV_\lambda$ is significant in North and South regions, this thesis infer the property of extreme rainfall in each model.

\subsubsection{Inference Under $PGEV_{\sigma}$ Model}

\begin{figure}
\centering \small 
{\includegraphics[width=15cm]{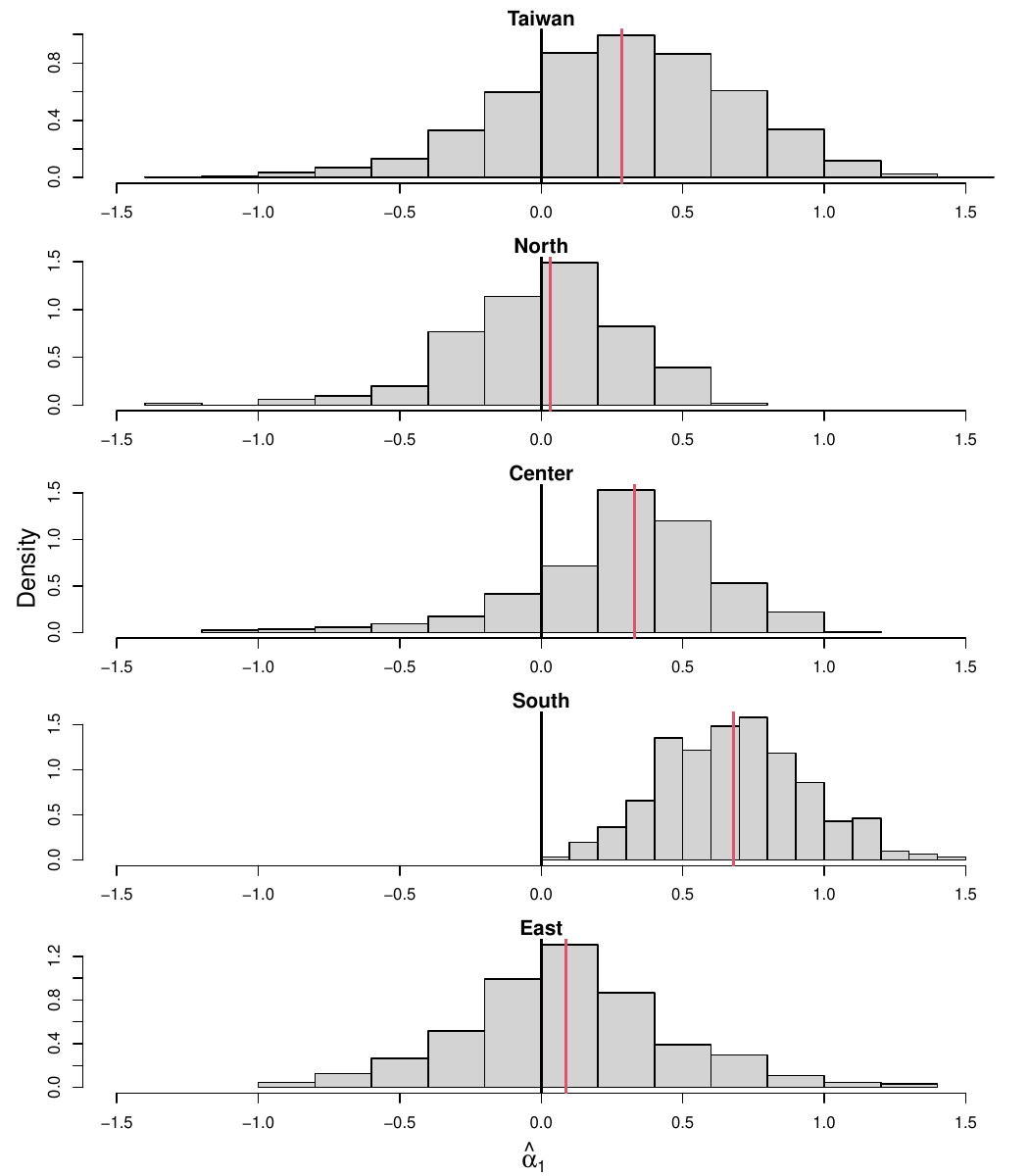}}
\caption[The histogram of $\hat{\alpha}_{1j}$ in $PGEV_\sigma$ models for different regions.]{The histogram of $\hat{\alpha}_{1j}$ in $PGEV_\sigma$ models in different regions. The dashed line means 0, and the solid line means the median of this distribution.} 
\label{Fig:pgev10 histogram region}
\end{figure}

Since $PGEV_{\sigma}$ is picked as the best model in Center and East regions, the main focus of this thesis is on the parameter $\hat{\alpha}_{1j}$, which represents the impact of temperature on the intensity of extreme rainfall events. To see the distribution of $\hat{\alpha}_{1j}$, Fig \ref{Fig:pgev10 histogram region} shows the histogram of $\hat{\alpha}_{1j}$ for the entire Taiwan region and in 4 subregions.
From Fig \ref{Fig:pgev10 histogram region}, the median of $\hat{\alpha}_{1j}$ (denoted as the solid red line) is positive in all regions. Also, more than half of the pixels are experiencing an increase in the intensity of extreme rainfall subject to temperature raise. Moreover, all pixels in the south have positive $\hat{\alpha}_{1j}$, indicating a substantial impact of temperature raise on extreme rainfall in this region. In contrast, the median in the north is near to 0, showing less evidence of the climate impact on extreme rainfall events.

To bring the location information into the $PGEV_\sigma$ model, and also to make plots spatially smoothed, a Gaussian Process (GP) is post-assumed to $\alpha_1(\bm{s}_j)\equiv \hat{\alpha}_{1j}$ to perform a high-resolution map of $\alpha_1(\bm{s})$ predict by 1km $\times$ 1km gridded pixels in Taiwan using kriging, displayed in Fig \ref{Fig:pgev10 map}. The detailed process of constructing the GP model is provided in Appendix I.
From Fig \ref{Fig:pgev10 map}, $\alpha_1(\bm{s})$ is larger in the southern. It means that the temperature has more impact on extreme rainfall events once the temperature becomes higher. In contrast, the Yilan area has smaller value of $\alpha_1(\bm{s})$, indicating less impact of temperature 
on extreme rainfalls events in this region.

\begin{figure}
\centering \small
\begin{tabular}{cc}
\subfigure[The Taiwan map of $\hat{\alpha}_{1j}$ in $PGEV_\sigma$ models]{\includegraphics[width=8cm]{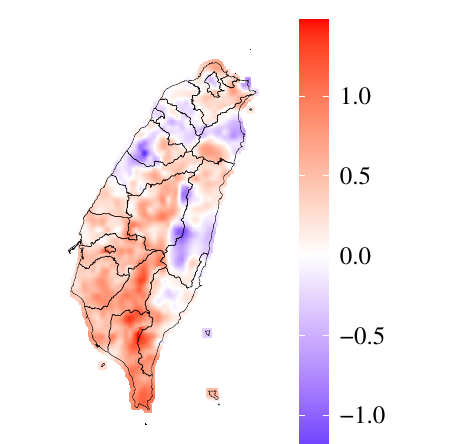}}&
\subfigure[The Taiwan map of t-value of $\hat{\alpha}_{1j}$ in $PGEV_\sigma$ models]{\includegraphics[width=6.5cm]{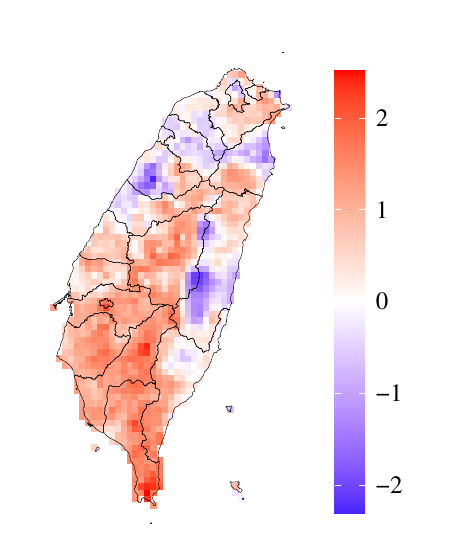}}
\end{tabular}
\caption[The plots for $\hat{\alpha}_{1j}$ in $PGEV_\sigma$ models]{Left plot is spatially smoothed map of $\alpha_1(\bm{s})$. Right plot is the t-value of $\alpha_1(\bm{s})$ in each pixel.} 
\label{Fig:pgev10 map}
\end{figure}


Finally, 4 hypothetic scenarios of temperature raise are investigated to further address the impact of climate change.
According to AR6 \citep{RN1}, the temperature changes are 1.5, 2, 3, and $4^{\circ}$C in near future relative to the period 1850-1900. Therefore, the temperature raises considered here are $\Delta_x=0.5, 1, 2$, and $3^{\circ}$C for the following analysis.

\begin{figure}[t]
\centering \small 
{\includegraphics[width=12cm]{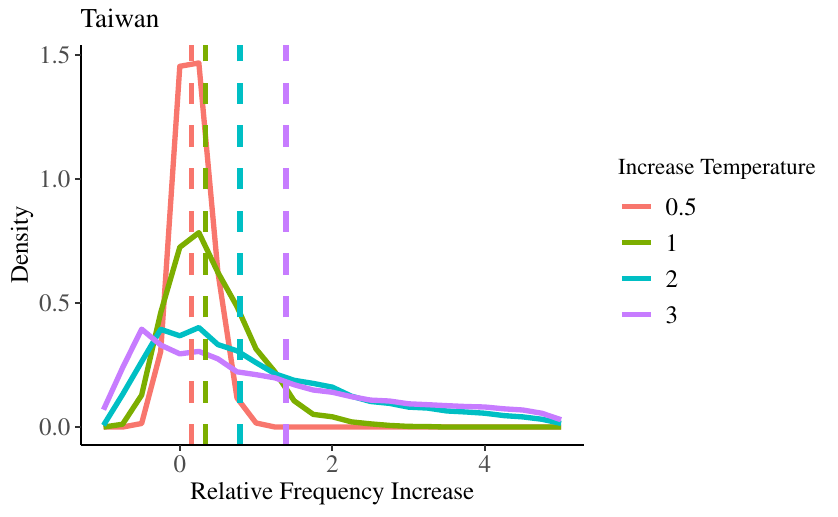}}
\caption[The intensity change in different temperature increase scenario]{Each curve is a distribution of relative intensity increase for each pixels under certain situation that temperature changes. The dashed line is the median of each curve. The x-axis is relative frequency change, and the y-axis is density of the distribution.} 
\label{Fig:relative frequency change}
\end{figure}

Fig \ref{Fig:relative frequency change} shows the distribution of 
$\Delta_\sigma$ among all pixels as a function of $\Delta_x$. 
The relative intensity $\Delta_\sigma$ is defined in Eq \eqref{eq:relative_scale}. 
The dotted line is the median of the individual distributions. 
Each line represents a temperature change scenario.
When $\Delta_x$ rises, the median of relative intensity $\Delta_\sigma$ goes higher. For example, the median of $\Delta_\sigma$ is around 16\% as $\Delta_x=0.5 ^\circ$C, and the median of $\Delta_\sigma$ is around 140\% as $\Delta_x=3 ^\circ$C.

\begin{figure}[t]
\centering \small 
{\includegraphics[width=12cm]{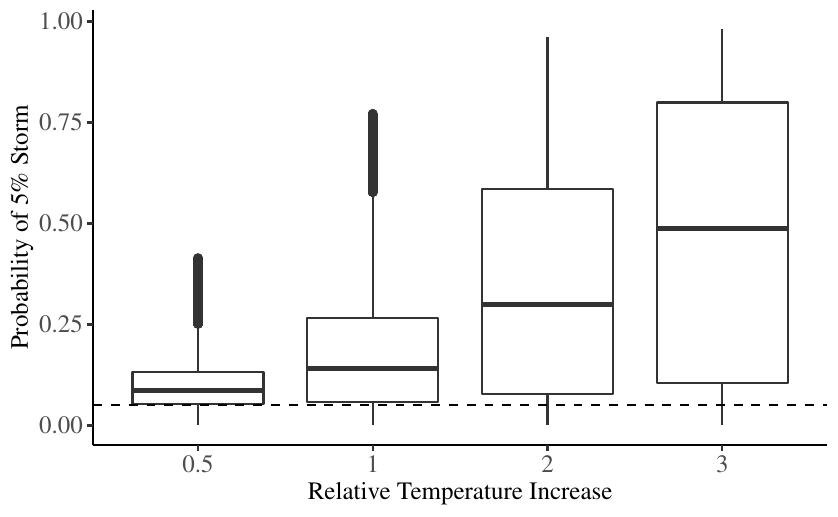}}
\caption[The probability of extreme storms with temperature increased]{The plot shows how much temperature rises, and the probability of extreme storm variation. The x-axis is the scenario of temperature increase, the y-axis is the probability of a 5\% storm under a certain scenario of temperature increase, and each whisker box is a distribution of probability that extreme storm happens when temperature changes, the dashed line is 5\% probability.} 
\label{Fig:extreme storm}
\end{figure}

Fig \ref{Fig:extreme storm} shows the probability of extreme storm variation defined in 
Eq \eqref{eq: relative_return_level_change} for $q=0.05$ under 4 temperature change scenarios. From Fig \ref{Fig:extreme storm}, the distribution of 5\% extreme storms becomes more often to happen as $\Delta_x$ increases. 
That is, if the temperature continues to increase in the future, the probability of happening extreme rainfall will become more than 5\%. In other words, originally the extreme rainfalls which happened every 20 years on average will occur more frequently when the temperature goes higher. It means in the future, the center and east of Taiwan will more likely to face extreme rainfalls than before, and the number of disasters followed by extreme rainfalls will also be increased, such as landslides and floods.

\subsubsection{Inference Under $PGEV_{\lambda}$ Model}

\begin{figure}
\centering \small 
{\includegraphics[width=15cm]{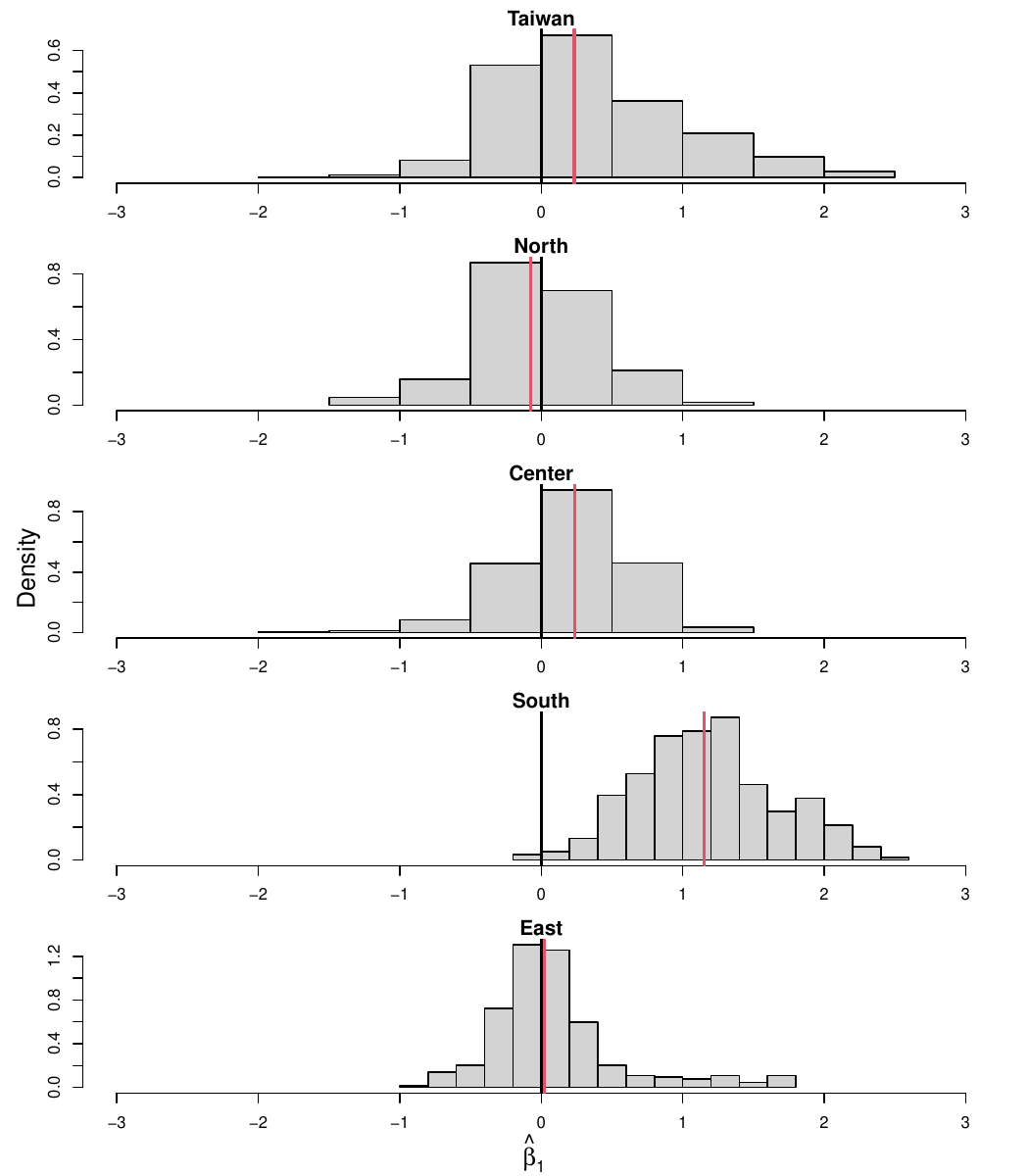}}
\caption[The histogram of $\hat{\beta}_{1j}$ in $PGEV_\lambda$ models for different regions.]{The histogram of $\hat{\beta}_{1j}$ in $PGEV_\lambda$ models in different regions. The dashed line means 0, and the solid line means the median of this distribution.} 
\label{Fig:pgev lambda histogram region}
\end{figure}

Since $PGEV_{\lambda}$ is picked as the best model in North and South regions, the main focus of this thesis is on the parameter $\hat{\beta}_{1j}$, which represents the impact of temperature on the frequency of extreme rainfall events. To see the distribution of $\hat{\beta}_{1j}$, Fig \ref{Fig:pgev lambda histogram region} shows the histogram of $\hat{\beta}_{1j}$ for the entire Taiwan region and in 4 subregions.
From Fig \ref{Fig:pgev lambda histogram region}, the median of $\hat{\beta}_{1j}$ (denoted as the solid red line) is positive in center and south regions. Moreover, most pixels in the south have positive $\hat{\beta}_{1j}$, indicating a substantial impact of temperature raise on extreme rainfall in this region. In contrast, the median in the north and east are near 0, showing less evidence of the climate impact on extreme rainfall events.

\begin{figure}
\centering \small
\begin{tabular}{cc}
\subfigure[The Taiwan map of $\hat{\beta}_{1j}$ in $PGEV_\lambda$ models]{\includegraphics[width=7cm]{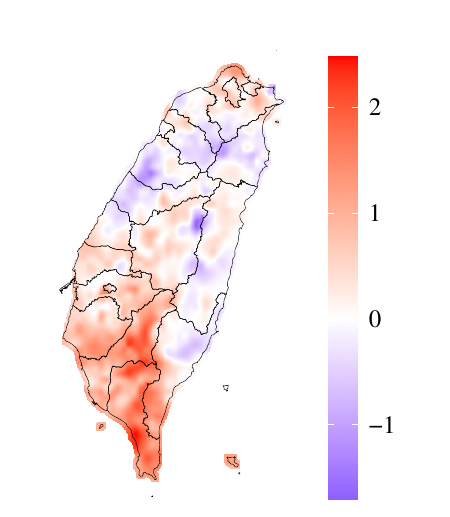}}&
\subfigure[The Taiwan map of t-value of $\hat{\beta}_{1j}$ in $PGEV_\lambda$ models]{\includegraphics[width=6.5cm]{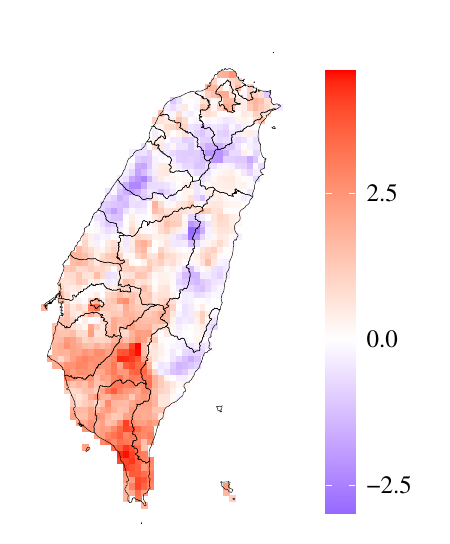}}
\end{tabular}
\caption[The plots for $\hat{\beta}_{1j}$ in $PGEV_\lambda$ models]{Left plot is spatially smoothed map of $\beta_1(\bm{s})$. Right plot is the t-value of $\beta_1(\bm{s})$ in each pixel.} 
\label{Fig:pgev lambda map}
\end{figure}

The same as Fig \ref{Fig:pgev10 map} map, a Gaussian Process (GP) is used to perform a high-resolution map of $\beta(\bm{s})$ in Fig \ref{Fig:pgev lambda map}. From Fig \ref{Fig:pgev lambda map}, $\beta(\bm{s})$ is larger in the southern. It means that the temperature has more impact on extreme rainfall events once the temperature becomes higher. In contrast, the Hualien area has smaller value of $\beta(\bm{s})$, indicating less impact of temperature on extreme rainfalls events in this region.

\begin{figure}[t]
\centering \small 
{\includegraphics[width=12cm]{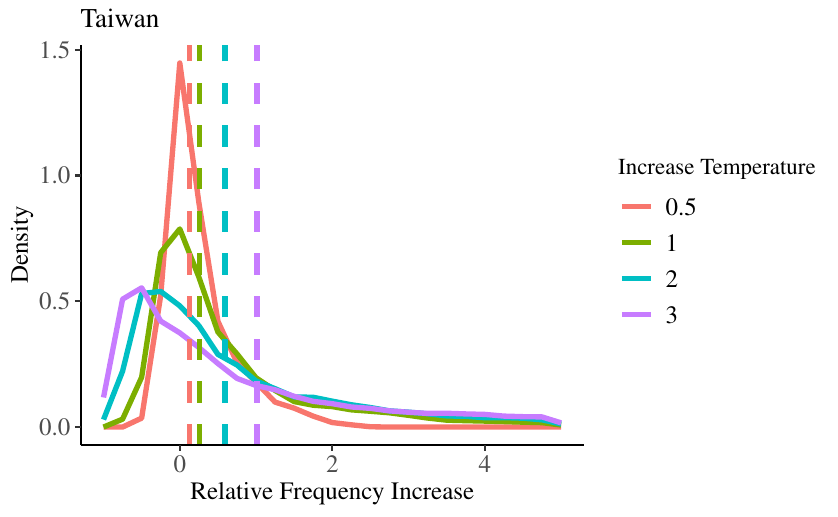}}
\caption[The frequency change in different temperature increase scenario]{Each curve is a distribution of relative frequency increase for each pixels under certain situation that temperature changes. The dashed line is the median of each curve. The x-axis is relative frequency change, and the y-axis is density of the distribution.} 
\label{Fig:relative frequency change lambda}
\end{figure}

Fig \ref{Fig:relative frequency change lambda} shows the distribution of $\Delta_\lambda$ among all pixels as a function of $\Delta_x$. The relative frequency $\Delta_\lambda$ is defined in Eq \eqref{eq:relative_intensity}. The dotted line is the median of the individual distributions. Each line represents a temperature change scenario.
When $\Delta_x$ rises, the median of relative frequency $\Delta_\lambda$ goes higher. For example, the median of $\Delta_\lambda$ is around 12\% as $\Delta_x=0.5 ^\circ$C, and the median of $\Delta_\lambda$ is 100\% as $\Delta_x=3 ^\circ$C.

\begin{figure}[t]
\centering \small 
{\includegraphics[width=12cm]{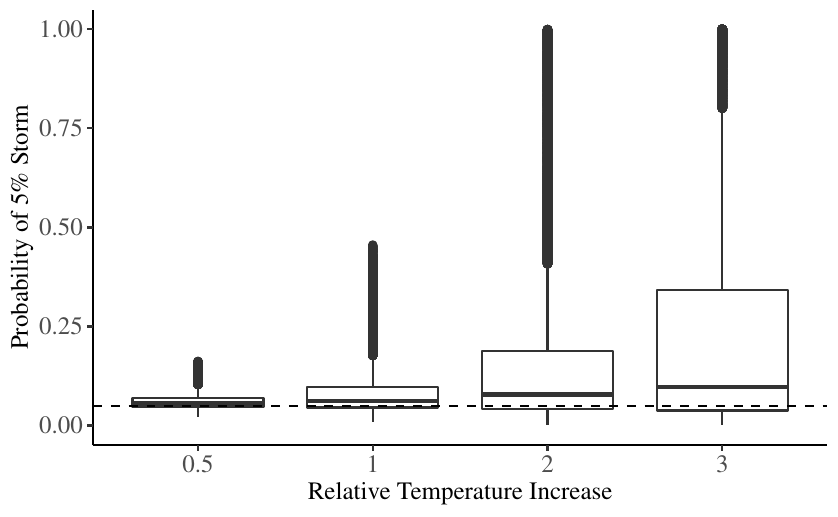}}
\caption[The probability of extreme storms with temperature increased]{The plot shows how much temperature rises, and the probability of extreme storm variation. The x-axis is the scenario of temperature increase, the y-axis is the probability of a 5\% storm under a certain scenario of temperature increase, and each whisker box is a distribution of probability that extreme storm happens when temperature changes, the dashed line is 5\% probability.} 
\label{Fig:extreme storm lambda}
\end{figure}

Fig \ref{Fig:extreme storm lambda} shows the probability of extreme storm variation defined in Eq \eqref{eq: relative_return_level_change} for $q=0.05$ under 4 temperature change scenarios. From Fig \ref{Fig:extreme storm lambda}, the distribution of 5\% extreme storms becomes more often to happen as $\Delta_x$ increases. 
That is, if the temperature continues to increase in the future, the probability of happening extreme rainfall will become more than 5\%. In other words, originally the extreme rainfalls which happened every 20 years on average will occur more frequently when the temperature goes higher. The results is the same as Fig \ref{Fig:extreme storm}.

\subsection{Hypothesis Testing on Models with Covariate}

Since Fig \ref{Fig:simulation MVN} shows that the confidence bands of $\Lambda_{\lambda}$, $\Lambda_{\sigma}$, and $\Lambda$ overlap each other, there are two more LRTs to confirm the relation between $PGEV_{\lambda}$, $PGEV_{\sigma}$, and $PGEV_{\lambda, \sigma}$. The two LRT for each pixel $j$ are:
 
\begin{enumerate}
\item[] {\bf Test a} \quad\quad $H_0$: $\bm{\theta}_j\in\Omega_\lambda$ vs $H_1$: $\bm{\theta}_j\in\Omega$,
\item[] {\bf Test b} \quad\quad $H_0$: $\bm{\theta}_j\in\Omega_\sigma$ vs $H_1$: $\bm{\theta}_j\in\Omega$.
\end{enumerate}

Accordingly, the LRT test statistics for Tests a-b and their null distribution under $H_0$ are given as follows respectively:

\begin{enumerate}
\item [] {\bf Test a}  
\begin{align}
\tilde{\Lambda}_\lambda=-2\left[\ell(\hat{\bm{\theta}}_j^{(\lambda)})-\ell(\hat{\bm{\theta}}_j)\right]\sim\chi^2_1,
\end{align}
\item [] {\bf Test b} 
\begin{align}
\tilde{\Lambda}_\sigma=-2\left[\ell(\hat{\bm{\theta}}_j^{(\sigma)})-\ell(\hat{\bm{\theta}}_j)\right]\sim\chi^2_1.
\end{align}
\end{enumerate}

\begin{figure}
\centering \small 
{\includegraphics[width=15cm]{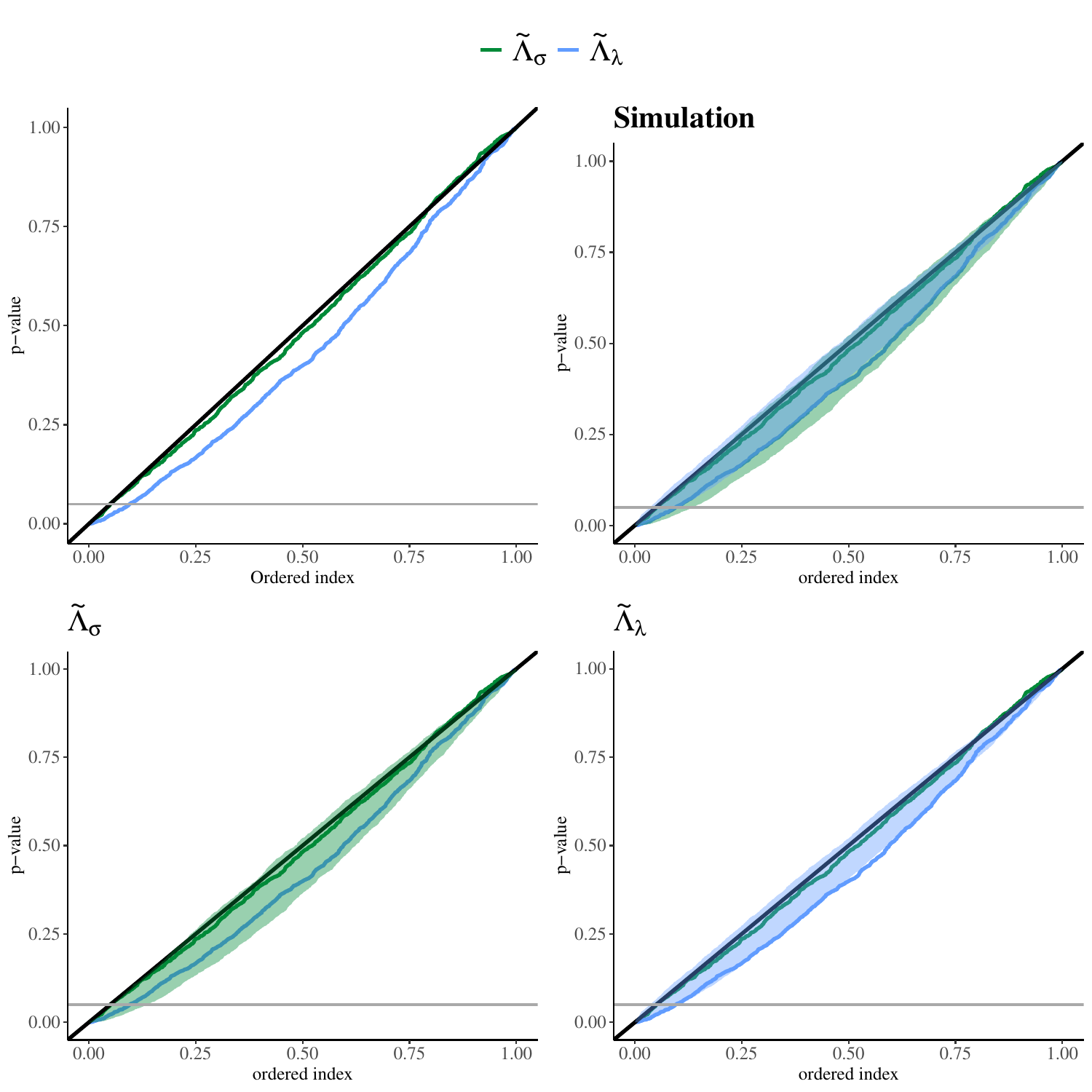}}
\caption[The PGEV with covariate p-value QQplot with confidence band]{The simulation result is the band in the picture above, and the curves represent Test a and b. The top left is the QQplot of p-values, the top right is QQplot with both confidence bands from all tests, and the others are the QQplot of p-values with confidence band for each test.} 
\label{Fig:advance_qqplot}
\end{figure}

The results of LRTs from Test a-b and corresponding confidence bands are shown in Fig \ref{Fig:advance_qqplot}. The curve represents the p-value of each test in each pixel $j$. All lines are close to the diagonal black line, and it means $PGEV_\lambda$ and $PGEV_\sigma$ are not significant compared with $PGEV_{\lambda,\sigma}$, especially all confidence bands include the diagonal line.  The results can be told from the confidence bands including the black diagonal line. Therefore, we can conclude that the models with covariates are not significant to each other.

\clearpage

\section{Discussion}

The purpose of this study was to examine the relationship between global warming and extreme rainfalls in Taiwan. To address this question, Taiwan rainfall and temperature data from TCCIP
were analyzed. 
Specifically, the PGEV model was used to fit the annual maximum of daily rainfall to infer the patterns of extreme values and possible impact due to climite change in temperature.

Results indicated that the intensity of extreme rainfall in Taiwan increases as the temperature rises, while the frequency remains constant.
However, the effects of global warming on the frequency and intensity of extreme rainfalls varied by region. In the north and south Taiwan regions, the frequency of extreme rainfalls changed with temperature, while in the central and east regions, the intensity of extreme rainfalls changed. The intensity of extreme rainfalls became higher as the temperature increased, according to the intensity plot. According to the 5\% return value distribution, extreme rainfalls will happen more frequently in the future due to temperature continuously increasing.

These findings have significant implications for water use and disaster prevention in Taiwan. Water in Taiwan mainly comes from the heavy rains brought by the mei-yu and typhoon season. If the temperature continues to rise in the future, Taiwan will face more frequent and intense extreme rainfall events, which could lead to serious floods and other types of disasters.

In 2009, typhoon Morakot caused catastrophic damage in Taiwan and is recorded as the peak in the functional box plots. The study also goes through the entire process of analysis without 2009 data, and the result is similar to the analysis with 2009.

For the post inference and visualization, Gaussian Process was used to spatially smooth the parameter estimates obtained from the fitted PGEV models and to construct the confidence bands for the p-value QQ plot. In particular, the latter task was implemented based on simulations using the Gaussian Process with Gaussian copula models. The confidence bands show that all tests comparing with models with and without covariates are significant, and the p-values from three tests are not significantly superior to others.

\clearpage

\bibliographystyle{apalike}
\addcontentsline{toc}{section}{References}
\bibliography{reference_2025}

\begin{thebibliography}{}

\bibitem[Akaike, 1974]{akaike1974new}
Akaike, H. (1974).
\newblock A new look at the statistical model identification.
\newblock {\em IEEE transactions on automatic control}, 19(6):716--723.

\bibitem[Blanchet and Davison, 2011]{blanchet2011spatial}
Blanchet, J. and Davison, A.~C. (2011).
\newblock Spatial modeling of extreme snow depth.
\newblock {\em The Annals of Applied Statistics}, pages 1699--1725.

\bibitem[Cleveland, 1979]{cleveland1979robust}
Cleveland, W.~S. (1979).
\newblock Robust locally weighted regression and smoothing scatterplots.
\newblock {\em Journal of the American statistical association},
  74(368):829--836.

\bibitem[Coles et~al., 2001]{coles2001introduction}
Coles, S., Bawa, J., Trenner, L., and Dorazio, P. (2001).
\newblock {\em An introduction to statistical modeling of extreme values},
  volume 208.
\newblock Springer.

\bibitem[Cotter and Dowd, 2006]{cotter2006extreme}
Cotter, J. and Dowd, K. (2006).
\newblock Extreme spectral risk measures: an application to futures
  clearinghouse margin requirements.
\newblock {\em Journal of Banking \& Finance}, 30(12):3469--3485.

\bibitem[Davison and Huser, 2015]{davison2015statistics}
Davison, A.~C. and Huser, R. (2015).
\newblock Statistics of extremes.
\newblock {\em Annual Review of Statistics and its Application}, 2:203--235.

\bibitem[Davison et~al., 2012]{davison2012statistical}
Davison, A.~C., Padoan, S.~A., and Ribatet, M. (2012).
\newblock Statistical modeling of spatial extremes.
\newblock {\em Statistical science}, 27(2):161--186.

\bibitem[Gkillas and Katsiampa, 2018]{gkillas2018application}
Gkillas, K. and Katsiampa, P. (2018).
\newblock An application of extreme value theory to cryptocurrencies.
\newblock {\em Economics Letters}, 164:109--111.

\bibitem[Henningsen and Toomet, 2011]{henningsen2011maxlik}
Henningsen, A. and Toomet, O. (2011).
\newblock maxlik: A package for maximum likelihood estimation in r.
\newblock {\em Computational Statistics}, 26(3):443--458.

\bibitem[Henny et~al., 2021]{henny2021extreme}
Henny, L., Thorncroft, C.~D., Hsu, H.-H., and Bosart, L.~F. (2021).
\newblock Extreme rainfall in taiwan: Seasonal statistics and trends.
\newblock {\em Journal of Climate}, 34(12):4711--4731.

\bibitem[Huser and Wadsworth, 2022]{huser2022advances}
Huser, R. and Wadsworth, J.~L. (2022).
\newblock Advances in statistical modeling of spatial extremes.
\newblock {\em Wiley Interdisciplinary Reviews: Computational Statistics},
  14(1):e1537.

\bibitem[IPCC, 2021]{RN1}
IPCC (2021).
\newblock {\em Climate Change 2021: The Physical Science Basis. Contribution of
  Working Group I to the Sixth Assessment Report of the Intergovernmental Panel
  on Climate Change}, volume In Press.
\newblock Cambridge University Press, Cambridge, United Kingdom and New York,
  NY, USA.

\bibitem[Nolde and Zhou, 2021]{nolde2021extreme}
Nolde, N. and Zhou, C. (2021).
\newblock Extreme value analysis for financial risk management.
\newblock {\em Annual Review of Statistics and Its Application}, 8:217--240.

\bibitem[Nortey et~al., 2015]{nortey2015extreme}
Nortey, E.~N., Asare, K., and Mettle, F.~O. (2015).
\newblock Extreme value modelling of ghana stock exchange index.
\newblock {\em SpringerPlus}, 4(1):1--17.

\bibitem[Olafsdottir et~al., 2021]{olafsdottir2021extreme}
Olafsdottir, H.~K., Rootz{\'e}n, H., and Bolin, D. (2021).
\newblock Extreme rainfall events in the northeastern united states become more
  frequent with rising temperatures, but their intensity distribution remains
  stable.
\newblock {\em Journal of Climate}, 34(22):8863--8877.

\bibitem[Projection and Platform, 2023]{TCCIP2023}
Projection, T. C.~C. and Platform, I. (2023).
\newblock {\em Gridded observation data}.

\bibitem[Sheraz et~al., 2021]{sheraz2021extreme}
Sheraz, M., Nasir, I., and Dedu, S. (2021).
\newblock Extreme value analysis and risk assessment: A case of pakistan stock
  market.
\newblock {\em Economic Computation \& Economic Cybernetics Studies \&
  Research}, 55(3).

\bibitem[Sun and Genton, 2011]{sun2011functional}
Sun, Y. and Genton, M.~G. (2011).
\newblock Functional boxplots.
\newblock {\em Journal of Computational and Graphical Statistics},
  20(2):316--334.

\bibitem[Tung et~al., 2016]{tung2016evaluating}
Tung, Y.-S., Cheng-Ta, C., Seung-Ki, M., and Lee-Yaw, L. (2016).
\newblock Evaluating extreme rainfall changes over taiwan using a standardized
  index.
\newblock {\em TAO: Terrestrial, Atmospheric and Oceanic Sciences}, 27(5):7.

\end{thebibliography}
\clearpage


\section*{Appendix I: Gaussian Process Models on Estimated Parameters}

The data used in this thesis are gridded data, and the PGEV model fitting is done at each pixel independently. As a result,  the parameter estimates $\hat{\bm{\theta}}_j$ obtained in Section 5.2 are not spatially smoothed. 
As a post inference, to bring the location information into estimation, the kriging method is used to smooth out the estimates for each parameter in space for a better visualization. 
In particular, 
the Gaussian process model with a Mat\'{e}rn covariance function is assumed on each parameter. The Mat\'{e}rn covariance function satisfies
\begin{align*}
C(h)=\sigma_M^2\cfrac{2^{1-\nu}}{\Gamma(\nu)}\left(\cfrac{\| h \|}{\rho}\right)^{\nu}K_\nu\left(\| h \|/\rho\right),
\end{align*}
\noindent
where $h\in\mathbb{R}$ is the distance, $K_\nu(\cdot)$ is the modified Bessel function of the second kind of order $\nu$, $\sigma_M^2$ is scale parameters, $\nu$ is shape parameters, and $\rho$ is range parameters. Since the repeated measurements are removed, the nugget effect here is set to zero.

\begin{figure}[t]
\centering \small 
{\includegraphics[width=12cm]{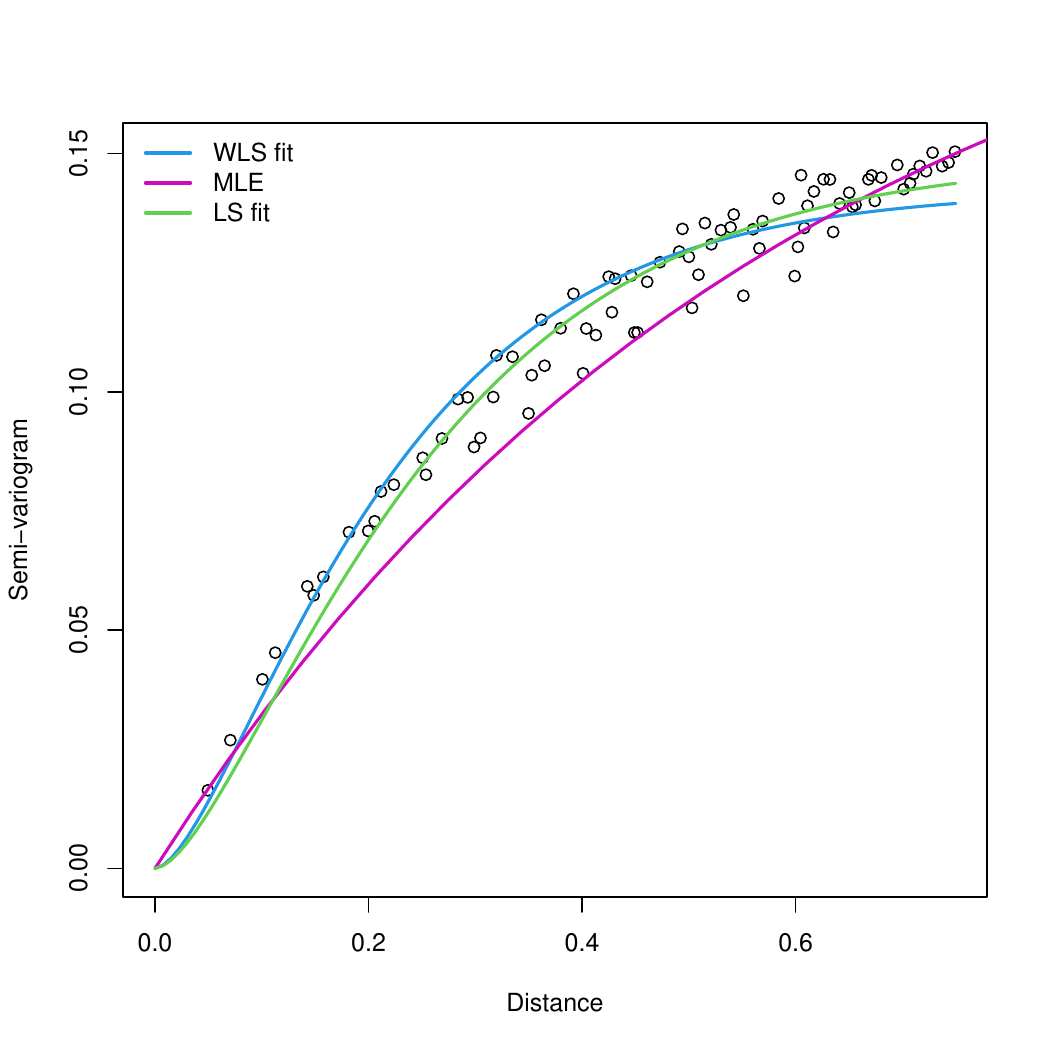}}
\caption[semi-variogram of $\alpha_1$ in $PGEV_\sigma$]{The sample semi-variogram of $\{\alpha_{1j}\}$ obtained from $PGEV_\sigma$ fitting. Mat\'{e}rn covariance model is fitted using LS method (green line), WLS method (Blue line), and MLE (purple line).} 
\label{Fig:semi-variogram}
\end{figure}

For illustration purpose, the kriging procedure for the parameter $\alpha_{1}$ is described as follows, which results in Fig \ref{Fig:pgev10 map}. Define $\alpha_1(\bm{s})$ be a Gaussian process with mean zero and covariance function $C(h)$, and the estimates $\alpha_{1j}$'s are treated as a realization of $\{\alpha_1(\bm{s}_j): j=1,...,1311\}$ respectively for each $j$, where $\bm{s}_j$'s represent the spatial locations on the grid. According to the semi-variogram fittings based on least squared or weighted least squared method and and MLE fitting, shown in Fig. \ref{Fig:semi-variogram}, this study chooses the fitting resulted from the semi-variogram fitting based on weighted least squared method to perform kriging shown in Fig \ref{Fig:pgev10 map}.  The fitting estimates are reported in Table \ref{tab:Matern Parameters Table}. The process is the same at $\beta_{1j}$ in Fig \ref{Fig:pgev lambda map}.

\begin{table}[b]
\caption{The parameters of Mat\'{e}rn($\sigma_M^2$, $\nu$, $\rho$) estimated using LS method, WLS method, and MLE.}
\small
\centering 
\begin{tabular}{lcccc}
\toprule[2pt]
method & scale parameters $\sigma_M^2$ & shape parameters $\nu$ & range parameters $\rho$ &  \\
\midrule
WLS   & 0.14         & 0.15         & 1.00  &  \\
LS       & 0.15          & 0.17         & 1.00  &  \\
MLE  & 0.21          & 0.60         & 0.50  &  \\
\bottomrule[2pt]
\end{tabular}
\label{tab:Matern Parameters Table}
\end{table}

\clearpage

\section*{Appendix II: Constructing Confidence Interval for QQ plot Using Simulation}

As in \citet{olafsdottir2021extreme}, the authors used simulation methods to construct the confidence interval for the QQ plot of p-values for PGEV model comparison. This thesis modifies their simulation method by considering dependence between parameters and spatial locations.
The simulation data are generated as follows: 
Suppose the original data follow $\bm{z}(\bm{s}_j)\sim F_c(\bm{z}(\bm{s}_j);\lambda_{j},\sigma_{j},\gamma_j)$,
\begin{enumerate}
\item original data $\bm{z}(\bm{s}_j)$ are transformed to Gaussian distribution $\bm{w}(\bm{s}_j)$ by the following formula.

\begin{align*}
\bm{w}(\bm{s}_j) = \Phi^{-1}(F_c(\bm{z}(\bm{s}_j)))\sim N(0,1),
\end{align*} 
\noindent
where $\Phi$ is CDF of standard normal distribution.

\item the Gaussian process models are fitted among transformed data. At this step, 61 Gaussian process models are bulit for each year ($t=1,...,61$), denoted as $M_t$. That is,

\begin{align*}
\{w_t(\bm{s}_j):j=1,\dots,1311\}\sim M_t.
\end{align*}

\item Gaussian Process models are sampled by bootstrap, denoted as $\tilde{M}_t$, and generate simulated data $\{\tilde{w}_t(\bm{s}_j):j=1,\dots,1311\}$ for each year $t$.
\begin{align*}
\{\tilde{w}_t(\bm{s}_j):j=1,\dots,1311\}\sim \tilde{M}_t.
\end{align*}

\item the simulated data $\bm{\tilde{w}}(\bm{s}_j)$ are transformed back to the original scale $\bm{\tilde{z}}(\bm{s}_j)$ with following,
\begin{align*}
\bm{\tilde{z}}(\bm{s}_j)\sim F^{-1}_c(\Phi(\bm{\tilde{w}}(\bm{s}_j))).
\end{align*}

\end{enumerate}
The whole simulation process repeat 100 times, including finding thresholds, fitting PGEV models, and getting the p-value with LRT.
The detailed procedures of the simulation is given in Algorithm \ref{alg:Simulation}.

After doing the above procedure 100 times, find the 0.95 quantile confidence interval for the QQ plot curve among the 100 simulations, and plot the simulationed confidence band in Fig \ref{Fig:simulation MVN}.

\begin{algorithm}
  \caption{Steps for Simulation}
  \label{alg:Simulation}
  \begin{algorithmic}[1]
  \FOR{$j\in[1,1311]$}
    \STATE Use CDF of PGEV in Eq \eqref{eq:PGEV} to transform original data $\bm{z}(\bm{s}_j)$ from PGEV into uniform distribution $\bm{u}(\bm{s}_j) = F_c(\bm{z}(\bm{s}_j)) \sim U(0,1)$ from Eq \eqref{eq:PGEV}.
    \STATE Use inverse CDF of a standard normal distribution to transform data from uniform distribution to standard Gaussian distribution $\bm{w}(\bm{s}_j) = \Phi^{-1}(\bm{u}(\bm{s}_j))\sim N(0,1)$.
    \ENDFOR
    \FOR{$t\in[1960,2020]$}
    \STATE Fit Gaussian Process $M_t$ with $\left\{w_t(\bm{s}_j):j=1,\dots,1311\right\}$ for each year $t$.
    \ENDFOR
    \STATE Bootstrap Gaussian Process models as $\tilde{M}_t$, and simulate the data $\left\{\tilde{w}_t(\bm{s}_j)\sim \tilde{M}_t:j=1,\dots,1311\right\}$ for each year $t$.
    \FOR{$j\in[1,1311]$}
    \STATE Use CDF of a standard normal distribution to transform simulated data into uniform distribution $\tilde{\bm{u}}(\bm{s}_j) = \Phi(\tilde{\bm{w}}(\bm{s}_j))\sim U(0,1)$
    \STATE Use inverse CDF of PGEV model to transfrom simulated data into PGEV distribution $\tilde{\bm{z}}(\bm{s}_j) = F^{-1}_c(\tilde{\bm{u}}(\bm{s}_j))\sim F(\tilde{\lambda_{c}}(t,\bm{s}_j),\tilde{\sigma_{c}}(t,\bm{s}_j),\tilde{\gamma}_j)$
    \STATE Fit GEV model in Eq \eqref{eq:GEV} with simulated data ignoring temperature and use Eq \eqref{eq:threshold} to get thresholds $\tilde{c}_j$.
    \STATE Fit PGEV four models in Eq \eqref{eq:PGEV} and use LRT to get the p-value of three different tests with covariates $x_t$ compared to the model without temperature.
    \ENDFOR
  \end{algorithmic}
\end{algorithm}

\clearpage

\section*{Appendix III: Comparison with Thresholds and Sample Quantiles}

According to Section 4.1 and Eq \eqref{eq:threshold}, the thresholds $c_j$ are the $p$ quantile of daily rainfall data in each pixel $j$. Since the threshold $c_j$ came from the PoT model, this study wants to show the relation between sample quantile and thresholds. However, the sample quantiles, defined as $Q_p$, are collected after declustering the data since PoT is assumed that data over the threshold are uncorrelated. 

Decluster is a technique to deal with the correlation between data over the threshold, such as extreme rainfall for a week. The process of decluster is to find the clusters of exceedances and choose the maximum of the cluster as the representative data for that rainfall event. For each pixel $j$, suppose $\delta$ is a cluster length, and let the first cluster start at the time $t$, then the rainfall of the cluster $z'_t$ modified as follows:

\begin{align*}
z'_{t}=
max\left\{z_i;i\in(t,\dots,t+\delta-1)\right\}.
\end{align*}

However, it is more common to decide run length $r$ rather than cluster length $\delta$. Run length defined as a cluster is still active until $r$ consecutive values are below the thresholds. In this study, $r$ is decided as 3.

\begin{figure}[t]
\centering \small 
{\includegraphics[width=12cm]{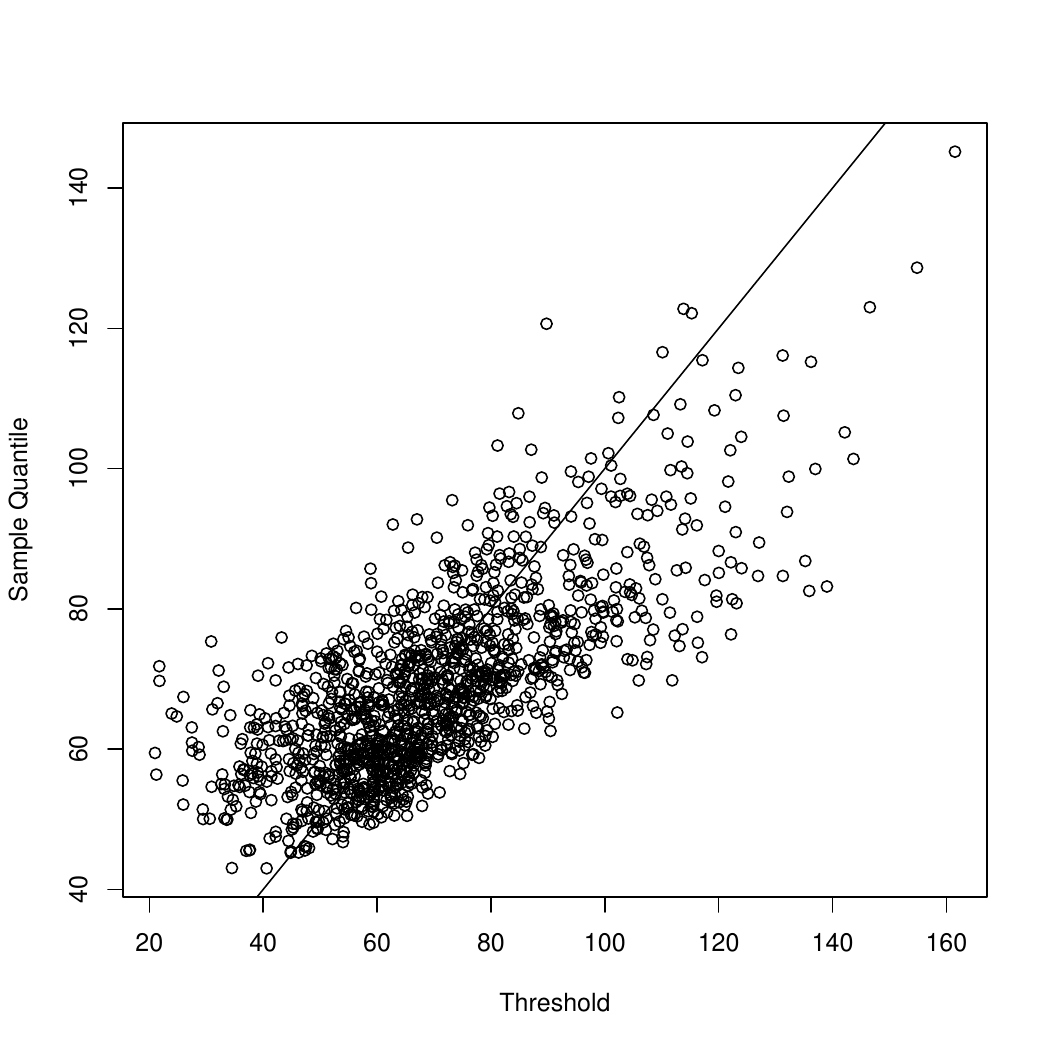}}
\caption[The comparsion between sample quantile and thresholds]{The x-axis is thresholds $c_j$ for each pixel $j$, the y-axis is the sample quantile $Q_p(\bm{s}_j)$ after decluster. The diagonal black line means $x=y$.} 
\label{Fig:threshold_quantile}
\end{figure} 

Fig \ref{Fig:threshold_quantile} shows the relation between sample quantiles and thresholds. Most points are around the diagonal line, which means most thresholds are near the sample quantiles.

\clearpage 

\end{document}